\documentclass[10pt,superscriptaddress,twocolumn,article]{revtex4}
\usepackage[format=plain,justification=justified,singlelinecheck=false,font=small,labelfont=bf,labelsep=space]{caption}
\usepackage{xr-hyper}
\usepackage[pdftex, pdfborder={0 0 0}, colorlinks=true, linkcolor=red,
urlcolor=blue]{hyperref}

\usepackage{caption}
\makeatletter
\newcommand*{\addFileDependency}[1]{
  \typeout{(#1)}
  \@addtofilelist{#1}
  \IfFileExists{#1}{}{\typeout{No file #1.}}
}
\makeatother

\newcommand*{\myexternaldocument}[1]{%
    \externaldocument{#1}%
    \addFileDependency{#1.tex}%
    \addFileDependency{#1.aux}%
}

\usepackage{fancyhdr,mathrsfs,times,mathptmx,balance,lastpage}
\usepackage{amsmath,amssymb,amsfonts,bm,graphicx,amsthm,cleveref}
\usepackage[toc,page]{appendix} \usepackage{color,subfigure}
\definecolor{red}{rgb}{0.75,0,0} \definecolor{blue}{rgb}{0,0,0.75}
\definecolor{green}{rgb}{0,0.5,0} \definecolor{orange}{rgb}{1,.34,.2}
\definecolor{purple}{RGB}{97,0,161} 
 \newcommand{\RR}{\mathbb{R}} 

\myexternaldocument{Supp_MultiControl_2021_02}

\begin{document}

\title{Structural underpinnings of control in multiplex networks}
\author{Pragya Srivastava} \affiliation{Department of
Bioengineering, University of Pennsylvania, Pennsylvania, PA 19104, USA}
\author{Peter J. Mucha} 
\affiliation{Department of Mathematics, University of North Carolina, Chapel Hill, NC 27599, USA}
\affiliation{Department of Applied Physical Sciences, University of North Carolina, Chapel Hill, NC 27599, USA}
\author{Emily Falk}
\affiliation{Annenberg School for Communication, University of Pennsylvania, 
Philadelphia, PA 19104, USA}
\author{Fabio Pasqualetti} \affiliation{Department of Mechanical
Engineering, University of California, Riverside, CA 92521, USA}
\author{Danielle S. Bassett} \affiliation{Department of Bioengineering,
University of Pennsylvania, Pennsylvania, PA 19104, USA}
\affiliation{Department of Physics \& Astronomy, University of Pennsylvania,
Pennsylvania, PA 19104, USA} \affiliation{Department of Electrical \& Systems
Engineering, Pennsylvania, PA 19104, USA} \affiliation{Department of Neurology,
Pennsylvania, PA 19104, USA} \affiliation{Department of Psychiatry,
Pennsylvania, PA 19104, USA} \affiliation{Santa Fe Institute, Santa Fe, NM 87501, USA}

\date{\today}

\begin{abstract}
To design control strategies that predictably manipulate a system's behaviour, it is first necessary to understand how the system's structure relates to its response. Many complex systems can be represented as multilayer networks whose response and control can be studied in the framework of network control theory. It remains unknown how a system's layered architecture dictates its control properties, particularly when control signals can only access the system through a single input layer. Here, we use the framework of linear control theory to probe the control properties of a duplex network with directed interlayer links. We determine the manner in which the structural properties of layers and relative interlayer arrangement together dictate the system's response. For this purpose, we calculate the exact expression of optimal control energy in terms of layer spectra and the relative alignment between the eigenmodes of the input layer and the deeper target layer. For a range of numerically constructed duplex networks, we then calculate the control properties of the two layers as a function of target-layer densities and layer topology. The alignment of layer eigenmodes emerges as an important parameter that sets the cost and routing of the optimal energy for control. We understand these results in a simplified limit of a single-mode approximation, and we build metrics to characterize the routing of optimal control energy through the eigenmodes of each layer. Our analytical and numerical results together provide insights into the relationships between structure and control in duplex networks. More generally, they serve as a platform for future work designing optimal \emph{interlayer} control strategies. 
\end{abstract}

\maketitle

\section{Introduction} 

The field of network science provides a powerful framework for studying the structure and organization of many complex systems including socioeconomic networks, traffic and communication networks, physical networks, and biological networks \cite{Gosak_bioNetsreview_2018, bassett2018on, LiaGranNetRev_2018, jackson_social_evol, jackson_socio_book}. Markedly useful and still an active area of research, the network representation of a complex system for a given research problem most often takes into account pairwise interactions of
only one type \cite{LeoAnn_complexreview}. However, in most complex systems, different types of interactions among system components can coexist and influence one another \cite{bianconimultilayer,kivela2014multilayer}. For example, in a transportation network people can move along bus, subway, or
train routes. In a socioeconomic network, the same set of individuals or institutions can interact socially as well as via financial transactions. Financial networks themselves can have 
multiple layers of interactions amongst their components \cite{bookstaber_multifinance, Wojcik_2018}. In a social network, 
the same set of users can interact in-person or via multiple social platforms \cite{Dickison_multisocial,atkisson_arxiv_multi}. Finally, in neuronal networks information can propagate along electrical and chemical synapses. The resultant structural picture of the system is thus that of a multilayer network where each component (node) can engage in multiple types of interactions with another component (node) across many temporal and spatial scales
\cite{Buldu_2017,Buldu_comment_2018,Muldoon2018comment, PilosofMultilayerEcology_2017, domenico2013mathematical, bentley2016multilayer, kidd2014unifying, Schmidt2018:MultiScale,bassett2011understanding,braun2018maps}.\\

Multilayered networks in true biological, physical, social, and technological realms do not exist in isolation, but instead impact \textemdash and are impacted by \textemdash  other systems around them through a variety of forces, influences, and causal
interactions \cite{Magnusson_1996, Bronfenbrenner_1979}. Yet, precisely how multilayer networks respond to perturbative signals, either beneficent or maleficent, remains far from understood. A promising route to progress is the utilization of linear systems theory, which provides a standard framework to understand the response of a complex system to both extrinsic and intrinsic perturbations \cite{kailath1980linear, kubo_toda_NEQ_linear}. The theory of linear control then utilizes the response of the system to determine its control properties and to inform the development of appropriate control strategies \cite{kailath1980linear}. An important part of characterizing system response is determining how a system's control properties depend upon its structure. While in monolayer networks the link between the structure and control properties has been investigated \cite{JasonKim_2018, Tang_2018},
this task becomes particularly challenging for multiplex networks. \\

Several recent studies have investigated the controllability of multiplex networks from various perspectives and have provided important initial insights. Using the minimum number of driver nodes $n_D$, that ensures the full controllability of \emph{directed} duplex networks, Posfai \emph{et al}.\ calculated the contribution of a given layer to the full controllability of a duplex network as a function of the relative time scales of activity update in the two layers \cite{posfai2016congtrollability}. In a complementary study, Menichetti \emph{et al}.\ investigated the role of degree correlations across layers on the minimum number of driver nodes necessary for control, particularly in settings where only a fixed set of nodes are control nodes in all layers \cite{menichetti2016control}. Moving beyond directed multiplex networks, Yuan \emph{et al}.\ obtained an exact expression for $n_D$ in terms of network size and the rank of the adjacency matrix associated with the duplex network \cite{Yuan_exact_control_multi}. Finally, Wang \emph{et al}.\ identified a trade-off between controllability and control energy for different patterns of interlayer connections in a duplex network \cite{Wang_multilayer_control_2017}. Although these studies have explored different aspects of the structure-control relationship, a generic framework that allows us to extract the dependence of multiplex control on layered architecture is currently lacking. \\

\begin{figure*}[ht]
\begin{center}
\includegraphics[scale=0.3]{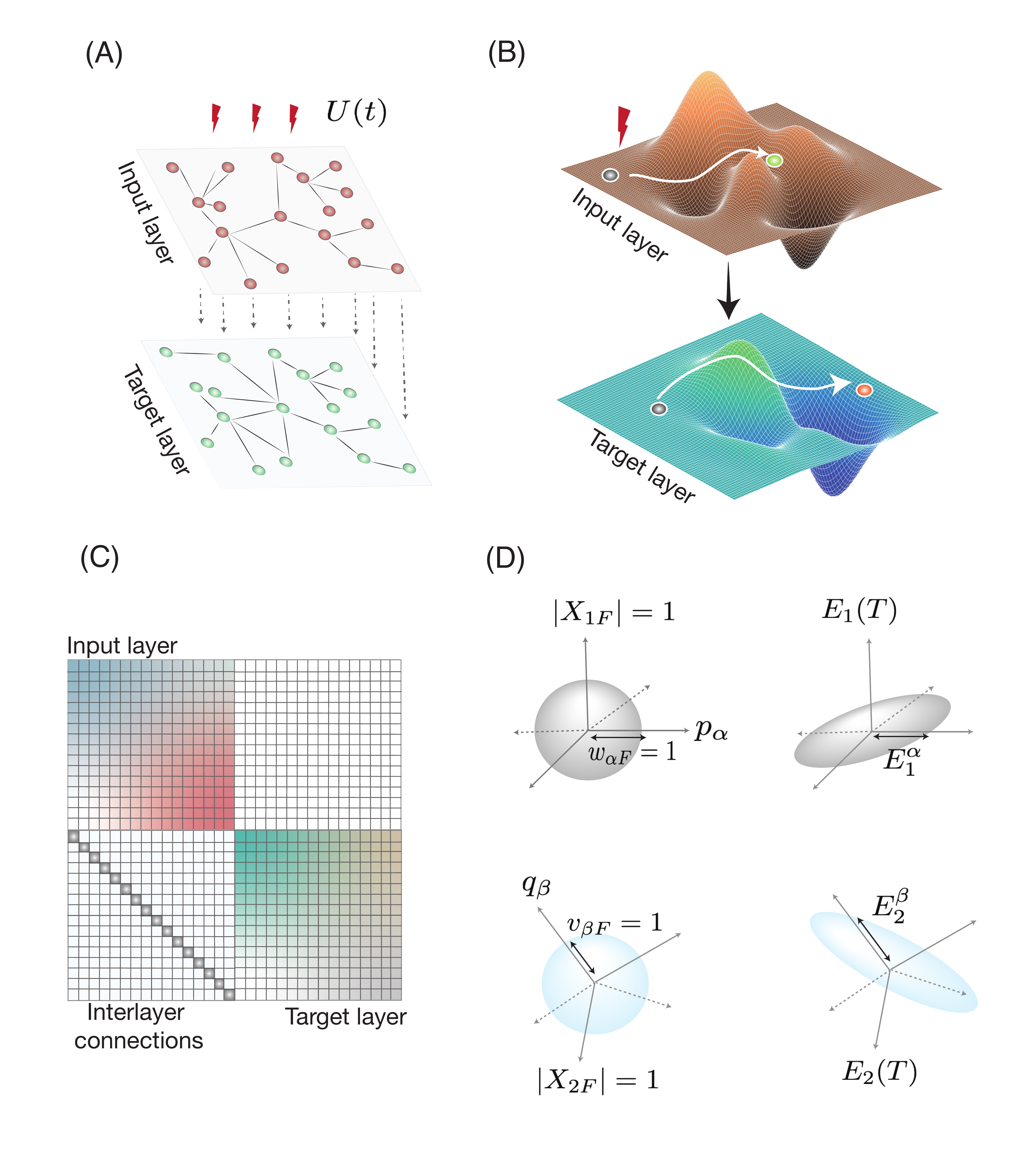}
\caption{\textbf{Control-theoretic framework for duplex networks.} (A)-(B) Control signals that are injected into the superficial, accessible input layer of the duplex network influence the state of the input layer, which in turn influences the state of the deep, inaccessible target layer via directed interlayer connections. (B) Stylized schematics of the energy landscapes of the input and the target layers on which the state trajectories are navigated by control signals. (C) The subdiagonal block of the supradjacency matrix for the duplex networks in this work has non-zero diagonal entries corresponding to the directed interlayer connections between replica nodes. The intra-layer networks have undirected connections resulting in a symmetric adjacency matrix. (D) Optimal energies for both layers are calculated corresponding to the navigation of their state along their eigendirections (denoted by $p_\alpha$ for the input layer and $q_\beta$ for the target layer) by a unit amount ($ w_{\alpha F} =1 $ for the input layer, and $v_{\beta F}=1$ for the target layer) from an initial state at the origin. In the eigenspace of each layer, final states ($X_{1F}$ for the input layer and $X_{2F} $ for the target layer) fall on the surface of the unit sphere, while the corresponding energies ($E_1^\alpha$ for the input layer and $E_2^\beta$ for the target layer) constitute the axes of an ellipsoid \cite{Wicks1988energyapproach}.
}
\label{fig:fig1}
\end{center}
\end{figure*}

Here, we investigate the relationship between structure and control in duplex networks, where we seek to drive the state of a deep, inaccessible target layer by injecting control signals into a superficial, accessible input layer (Fig.~\ref{fig:fig1}(A)). 
Such control requirements could arise in real multilayered systems
where external signals can manipulate a specific type of interaction due to constraints or convenience. For instance, in the context of social networks an objective could be to modify the behaviour of individuals in their professional environment (professional layer) by providing resources of support in their informal social environment. In multilayered socioeconomic networks, reduction of vulnerability to financial crises in a network of financial institutions can be achieved by designing optimal intervention strategies by the government and public. In multilayered brain networks, an important control objective could be to manipulate inter-regional connectivity (i.e. the structural connectivity matrix) by targeting the activity of specific brain-regions.\\ 

Having specified such a control configuration, we ask how the structural properties of
individual layers (eigenspectra) and their relative arrangement (alignment between layer eigenmodes) together impact control. Using the framework of linear control theory in a layer-diagonalized representation, we determine the optimal energy required to drive the system to a given final state, and show analytically how it depends upon these two structural parameters. Using numerical calculations, we then estimate the optimal control energy for duplex networks whose layers are Erd\H{o}s-R\'{e}nyi (ER), Watts-Strogatz (WS), Barab\'{a}si-Albert (BA), or Random-Geometric (RG) graphs. 
We calculate the average and maximum control energy of the target layer as a function of the layer density and topology. We show that the alignment between the eigenmodes of the two
layers determines the cost of controlling specific eigenmodes of the target layer. Moreover, the angle of alignment determines the number of eigenmodes that are excited by the optimal control energy. More specifically, we demonstrate the routing of optimal control energy through the eigenmodes of the input layer in a manner that depends systematically on their alignment with the
target eigenmode. The approach and results presented in this work can be utilized to examine the interlayer relationships in complex multilayer systems, identify and predict the easy versus difficult control directions, and analyse the physical paths along which control is exercised. \\

The remainder of this paper is structured as follows. In Section \ref{sec:formulation}, we formulate the problem of controlling a duplex with linear time-invariant (LTI) dynamics and outline our approach. In Section \ref{sec:numerics}, we present the numerical results on the structural dependence of the average control energy that emphasize the role of target-layer topology, interlayer connections, and alignment. Then, in Section~\ref{sec:onemode} we deepen our understanding of these results in the simplified limit of a one-mode approximation. In Section~\ref{sec:pathdep_control}, we use the insights from our results to build metrics to identify the paths along which optimal control signal is distributed. Finally, in Section~\ref{sec:conclusion&future}, we conclude our investigation and discuss future directions.

\section{Formulation} \label{sec:formulation}
We consider a duplex network in which control signals are injected into the first layer, also called the `input' layer (Fig.~\ref{fig:fig1}(A)). Practically, we think of the input layer as a superficial layer in a real system which is accessible to external perturbations. Since different layers of a multiplex network represent different types of interactions between system-components, this control setup mimics the case when only one type of interaction can be directly affected by the external perturbations. Any changes in the state of the first layer flow to the second `target' layer via directed interlayer links (Fig.~\ref{fig:fig1}(A)-(B)). For simplicity, we assume that the intralayer connectivity matrices are undirected. The supra-adjacency matrix $M$ for such a duplex network has the following form \cite{kivela2014multilayer}: 

\begin{align}
 M = \left(\begin{array}{cc} A^{(1)} & 0_N  \\ K & A^{(2)}
\end{array} \right), \label{duplex} 
\end{align} 

\noindent where $A^{(1)} \in \RR^{N \times N}$, $A^{(2)} \in \RR^{N \times N}$, and $K \in \RR^{N \times N}$ (Fig.~\ref{fig:fig1}(C)). Here, $A^{(1)}$ and $A^{(2)}$ denote the \emph{intralayer} adjacency matrices of the two layers, and the matrix $K$ encodes the interlayer connections. In a duplex topology, each layer consists of the same set of nodes. Moreover, the interayer connections exist only between 
the same nodes; that is, a given node in a layer connects to its `replica' node in a different layer. Thus, the interlayer connectivity 
matrix $K$ assumes the form $K = \kappa I_N$, where $\kappa$ denotes the weight of interlayer connections, and $I_N \in \RR^{N \times N}$ is the identity matrix in $N-$ dimensions. Finally, $0_N$ denotes an $N-$ dimensional square matrix with zero entries. 

The control signal $u(t)$ is
injected into the first layer of this duplex via an input matrix $B$. We model the evolution of the network's state by a linear time-invariant dynamics in continuous time: 

\begin{align}
\frac{d X}{d t}= M X(t) + B u(t),
\label{eq:CT-LTI} \end{align}

\noindent where $X(t) \in \mathbb{R}^{2 N \times 1} $ is a vector that denotes the state of nodes in all network layers at time $t$. We assume that the control signal is injected into all of the nodes of the input layer, and thus $B$ is $B \in R^{2 N\times N}$ and $ B = \left(I_N \,\, 0_N\right)^T $. Clearly, $u(t) \in \RR^{N \times 1}$. Finally, the cost of control is measured by the control energy $E$, which is defined as: 

\begin{equation} 
E = \int_0^{t_F} u^T(t) u(t) d t ,
\end{equation}

\noindent where $t_F$ is the time horizon of the control, and the superscript $T$ denotes the matrix transpose.\\

In order to explicitly link the structure of a duplex network to the cost of control, we first express state equations in a layer-diagonalized representation. Specifically, we first separate the state vector $X$ in $X^{(1)} \in \RR^{N \times 1} $ and $X^{(2)} \in \RR^{N \times 1}$, where the former represents the state of the nodes in the input layer and the latter represents the state of the nodes in the target layer. Then, the LTI dynamics in continuous time (Eq.~\eqref{eq:CT-LTI}) can be written as:
    
\begin{subequations} 
\begin{align} 
\dot{X}^{(1)} &= A^{(1)} X^{(1)} + u
\label{eq:X1-dyn}\\
 \dot{X}^{(2)} &= K A^{(1)} + A^{(2)} X^{(2)} .
\label{eq:X2-dyn} 
\end{align}
\end{subequations}

\noindent We diagonalize the real-symmetric matrices $A^{(1)}$ and $A^{(2)}$ as $A^{(1)} = P D_1 P^T$ and $A^{(2)} = Q D_2 Q^T$, where $Diag(D_1) = \{ \xi_\alpha \vert \alpha \in \{1,2, \ldots, N \} \}$ and $ Diag(D_2) = \{ \mu_\alpha \vert \alpha \in \{1,2, \ldots, N \}\}$. In addition, the matrices $P$ and $Q$ have the form 
$P = \left[\mathbf{p}_1, \mathbf{p_2}, \ldots, \mathbf{p}_\alpha, \ldots, \mathbf{p}_N \right]$, and 
$Q = \left[\mathbf{q}_1, \mathbf{q_2}, \ldots, \mathbf{q}_\alpha, \ldots, \mathbf{q}_N \right]$, such that
$\mathbf{p}_\alpha$ and $\mathbf{q}_\alpha$ are the $\alpha^{\rm{th}}$ eigenvectors of $P$ and $Q$, respectively. Using the above decomposition, we next write, $ X^{(1)} = \underset{\alpha = 1}{ \Sigma} w_\alpha \mathbf{p}_\alpha$ and $X^{(2)} = \underset{\alpha =1}{\Sigma} v_\alpha \mathbf{q}_\alpha$. Rewriting 
Eqs.~\eqref{eq:X1-dyn} and \eqref{eq:X2-dyn}, we obtain: 

\begin{subequations}
 \begin{align} \dot{w}_\alpha &= \xi_\alpha w_\alpha +
 U_\alpha 
\label{eq:L1-LTI} \\
\dot{v}_\alpha & =  \mu_\alpha v_\alpha + \kappa
\underset{j}{\Sigma} w_\beta \cos \theta_{\alpha \beta},
\label{eq:L2-LTI-angle}
\end{align} 
\end{subequations}
\noindent
where we have used the fact that $u(t) = \Sigma U_\alpha \mathbf{p}_\alpha $, and that $K = \underset{\beta}{\Sigma} \mathbf{q}_\beta \mathbf{k}_\beta$. For the duplex network 
considered here, $K = \kappa I_N$ leading to $\mathbf{k}_\beta = \mathbf{q}_\beta^T$. This,
in turn, leads to the 
identification of $\theta_{\alpha \beta} = \arccos ( \mathbf{q}_\beta^T \mathbf{p}_\beta)$ as the angle between $\mathbf{q}_\alpha$ and $\mathbf{p}_\beta$. 
These decompositions correspond to the projection of $u(t)$ and the matrix $K$ along the eigenvectors of the input layer and the target layer, respectively. The resulting
Eqs.~\eqref{eq:L1-LTI} and \eqref{eq:L2-LTI-angle} have an explicit dependence on the eigenvalues of each layer as well as on the angles $\theta_{\alpha \beta}$ that characterize the alignment between the eigenmodes of the two layers. The state equations Eqs.~\eqref{eq:L1-LTI} and \eqref{eq:L2-LTI-angle} are obtained using the transformation which diagonalises the adjacency matrix of each layer, which we refer to as 
the `layer-diagonalised' representation of the state equations.\\

In a controllable system, there could be multiple control inputs that navigate the system between a pair of states. The task then becomes to find the minimum control energy that is required to control the system from an initial state to a desired final state. Mathematically, this minimization problem is expressed
as: 

\begin{align}
\underset{U(t)}{{\rm{minimize}}} \quad E & = \int_0^{t_F} U^T(t) U(t) d t , \nonumber\\
\text{subject to }  \dot{w}_\alpha &= \xi_\alpha w_\alpha + \mathbf{b}_\alpha
\cdot U \nonumber \\
\dot{v}_\alpha &= \mu_\alpha v_\alpha + \kappa
\underset{j}{\Sigma} w_\beta \cos \theta_{\alpha \beta} .
\label{eq:optim_statement} \end{align}

\noindent This problem can be solved analytically using the standard framework
of optimal control theory \cite{lewis_optimal_control_2012}. We apply this
framework to the problem of Eq.~\eqref{eq:optim_statement} and provide the
exact solutions for the state variables $\{ \{ w_\alpha \}, \{v_\alpha \} \}$
and the optimal control input $u_\alpha = \mathbf{b}_\alpha \cdot U$ in closed
form (see the supplementary section {\color{red} \bf S1}). \\ 

We note that our solutions to the minimization problem display a complex dependence on the two structural parameters of interest: the eigenvalues of each layer and the alignment of the eigenmodes of the two layers. This dependence renders our solutions rather opaque, and it remains difficult to extract simple insights into the structural determinants of control. To make progress, we therefore now turn to a numerical computation of the optimal control energy for duplex networks with layers constructed from canonical graph models. After this numerical assessment, we work in a single-mode limit to more deeply understand our results.

\section{Structure-control relationships in duplex networks: A numerical investigation}
\label{sec:numerics}

\begin{figure*}[ht]
\centering
\includegraphics[scale=0.2]{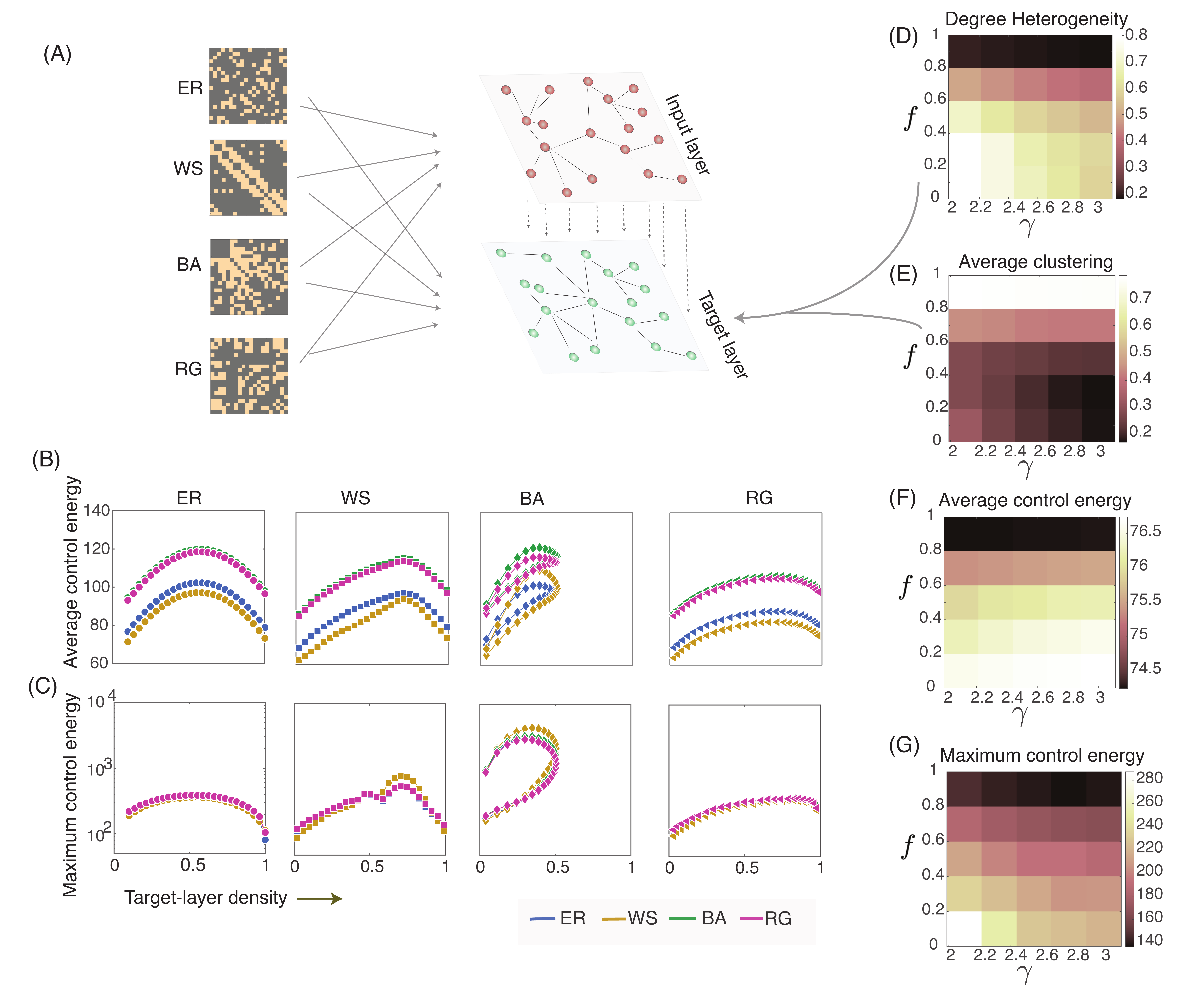}
\caption{\textbf{Control properties of simulated duplex networks.} (A) Schematic of the process by which we construct duplex networks. Each layer belongs to one of the following graph families; Erd\H{o}s-R\'{e}nyi (ER), Watts-Strogatz (WS), Barab\'{a}si-Albert (BA), and Random Geometric (RG). An example network from each of these graph families is shown on the left. (B) The plots of average control energy and (C) the maximum control energy of the target layer as a function of the target-layer density for all graph families studied. Each column corresponds to the topology of the target layer. Throughout 
this paper, we use differently-colored curves to indicate the topology of the input layer, and we use different markers to indicate the topology of the target layer. The average control energy is lowest for the duplex network with a WS network as the input layer and an RG network as the target layer. The loops in the plots corresponding to 
a BA network in the target layer are indicative of a non-monotonic dependence
of network density on the generating parameter (see supplementary
section and Fig. {\color{red} \bf S3}) (D)-(E). 
To further understand our results in graph model families, we build a set of hierarchically modular (HM) networks as a function of two generating parameters: $\gamma$, which determines the degree heterogeneity, and $f$, which determines the modularity. Here we show heatmaps of degree heterogeneity and average clustering for the HM networks. (F)-(G) The heatmaps of average and maximum optimal control energies of the target layer in duplex networks constructed from HM networks.
}
\label{fig:fig2}
\end{figure*}

In this section, we present our numerical results on the control properties of a duplex network as a function of the properties of its layers and interlayer alignment. In the previous section, we determined the dependence of control energy on: (i) the two sets of eigenvalues corresponding to the two layers of the duplex, and (ii) the angles characterizing the alignment between the eigenvectors of the two layers. Network properties that directly influence the eigenvalues are the density of the network and the topology of the network. To simulate the variations in topology, we construct duplex networks where each
layer belongs to one of the following four graph families: Erd\H{o}s-R\'{e}nyi (ER), Watts-Strogatz (WS),
Barab\'{a}si-Albert (BA), and Random-Geometric (RG). Networks in these four
families differ in their topology, exhibiting distinct degree heterogeneity and
clustering coefficient \cite{Albert-Barabasi_1999, Albert-Barabasi_2002_RMP,
Penrose2004_RGG, watts_strogatz_1998} (see the supplementary section
and Fig. {\color{red}\bf S2}); moreover, the RG
family is  spatially embedded while the other 3 are not. In addition, several
real systems can be represented by networks from these four families, making
their study germane to a generic understanding biological, physical, and social
systems \cite{Albert_SFBiol_2005, Keller_2005_SFessay,
Telesford_2011_WSubiquity, Penrose2004_RGG}. To simulate variations in network
density,
we changed the density of the target layer while holding the density of the
input layer fixed (see the supplementary section and Fig.~{\color{red}\bf S3}). \\

To explicitly relate the control properties of the duplex networks to their structure, we calculate the optimal control energy required to drive the system from the origin to a set of pre-specified final states. We chose the final states to be those that have unit norm and that are aligned with the eigenvectors of the target layer (Fig.~\ref{fig:fig1}(D)). Control energies thus calculated define the intersections of the control energy surface with the eigendirections of each layer. Intuitively, these control energies characterize the difficulty of navigating the state of a layer along a given eigendirection. We use the quantity $E_\ell^{\alpha}$ to denote the control energy of navigating the state of the $\ell^{\rm{th}}$ layer along the $\alpha^{\rm{th}}$ eigendirection by a unit amount. Then, a measure of the average controllability of a given layer can be defined as

\begin{align}
\bar{E}_\ell = \underset{\alpha}{\Sigma} E_\ell^\alpha, \quad \ell \in \{1,2 \} .
\label{aver_E}
\end{align}

\noindent Similarly, the maximum control energy corresponds to the eigendirection along which it is most difficult to control the system. To extract this direction, we define:

\begin{align} 
E_\ell^{\rm{max}} = \rm{max}\{E_\ell^{\alpha}: \alpha \in \{1,2,\ldots,N \}, \ell \in \{1,2 \} \}.
\label{max_E}
\end{align} 

\noindent For a given duplex network, the $E_\ell^\alpha$ quantify the difficulty of navigating the state of the $l^{\rm{th}}$ layer along its $\alpha^{\rm{th}}$ eigenvector, thus giving access to the complete spectrum of control properties. We calculate $E_\ell^\alpha$ for all duplex networks
constructed from the same four graph families studied earlier (see the supplementary Section 
and Fig.~{\color{red}\bf S4}). Since we primarily focus our attention on controlling a layer by applying the control in a different layer, in the following section we discuss the control energies only for the target layer. The following subsections report our results first for the eigenvalues of each layer and then for the alignment between the eigenmodes of the two layers.

\subsection{Density- and topology-dependence of the target-layer control} 
\label{subsec:topo_dep}

To calculate the control energies of the target layer, we create sixteen different duplex networks by constructing each layer from the 4 graph families (Fig.~\ref{fig:fig2}(A)). We normalize the supra-adjacency matrix of each duplex by the largest
eigenvalue $\xi_{\rm{max}}$ of the first layer, which enables us to consistently compare the results obtained for the different duplex networks. Since each eigenvalue has the dimensions of a rate, intuitively this normalization sets the unit of time as the inverse of the largest eigenvalue of the input layer. We set the strength of the interlayer connections, $\kappa$, to be the same as the strength of the intralayer connections in the \emph{unnormalized} network. Thus in normalized duplex networks, the strength of the interlayer connections is set to
$\kappa/\xi_{\rm{max}}$. We return to the dependence of control 
energy on $\kappa$ later in this paper (also see the supplementary section
and Fig.~{\color{red}\bf S5}). Finally, we calculate the quantities $\bar{E}_2$ and $E^{\rm{max}}_2$ as a function of the density of the target layer, for 40 realizations of each duplex network. \\

We find that the average control energy $\bar{E}_2$ depends upon the density as well as the topology of the input and target layers (Fig.~\ref{fig:fig2}(B)). For all duplex networks, the dependence of $\bar{E}_2$ on the target layer density is non-monotonic; consistently, the intermediate densities are the hardest to control. Further, $\bar{E}_2$ is lowest when the topology of the target layer is that of an RG graph, for all the duplex networks and for all density values. For a fixed topology of the target layer, $\bar{E}_2$ is lowest when the topology of the input layer is that of a WS network. Thus, a duplex network with a WS network in the input layer and an RG network in the target layer has the lowest average control energy. \\

Next, we find that the maximum control energy $E^{\rm{max}}_2$ also depends upon the density of the target layer (Fig.~\ref{fig:fig2}(C)). For all duplex networks, the maximum control energy is highest in the intermediate range of densities. The topology of the input layer has a weak effect on $E^{\rm{max}}_2$, when the target layer has the topology of an ER or an RG network. However if the target layer is a WS network or a BA network -- and if the input layer is a WS network--, then the $E^{\rm{max}}_2$ is quite high. This effect is most striking when the target layer has the topology of a BA network.
One possible reason for this behavior is that BA networks are characterized by a power-law degree distribution, implying 
the existence of many nodes that have either much higher or much lower degree than expected in an ER graph. Indeed, we show a similarity between the degree heterogeneity of the target layer and $E_2^{\rm{max}}$ in the next subsection. Intuitively, a larger number of sparsely connected nodes could contribute to the eigendirections that are much harder to control in BA
networks. \\ 

To evaluate the robustness of our findings, we performed several additional
sensitivity analyses. First, we note that in Fig.~\ref{fig:fig2}(B-C) the
average and maximum control energy are plotted for a fixed density of the input
layer. We verify that our results remain unchanged for different densities of
the input layer (see the supplementary section and Fig.~{\color{red} \bf S6}). 
Additionally, similar plots for $\bar{E}_1$ and $E^{\rm{max}}$ can be made
(see the supplementary section and Fig.~{\color{red} \bf S7}). Note that, for
the \emph{unnormalized} interlayer connection strength $\kappa =1$, we find
that the optimal control energies to control the target layer are much larger
than that of the input layer. Varying the strength of $\kappa$ changes the
relative values of the optimal control energy for input and target layers: a
large value of $\kappa$ leads to a faster propagation of control signals
between the two layers, thus lowering the control energies (see the
supplementary section and Fig.~{\color{red} \bf S5}). Note that we explain this dependence of
control energy on $\kappa$ more formally in the single-mode limit in Section
\ref{sec:onemode} of this main text. \\

In sum, unlike the input layer in which control signals are injected directly, the effect of control signals on the target layer is mediated by the input layer and the interlayer connections. This makes the strength of interlayer connections and the arrangement of the target layer relative to the input layer important parameters in determining the cost of controlling the target layer. In this section, we characterized global metrics of controllability, namely, the average and the maximum control energy of the target layer, as a function of
the density of the target layer and the topologies of the two layers. However, the trends we observe (Fig. \ref{fig:fig2}(B-C)) are due to the compound effects of network density and topology. In the next section, we separate the role of network topology from that of the network density. In particular, we focus on the role of heterogeneity in the target layer on its control properties. To do this, we construct duplex networks whose target layers continuously span a range of degree heterogeneity and clustering coefficient. This approach allows us to explicitly link the two topological properties of the target layer to the cost of their control.

\subsection{Control properties of a heterogeneous target layer}
\label{subsec:HMnets}

Many real-world networks exhibit topological properties that differentiate them from Erd\H{o}s-R\'{e}nyi networks. Examples include heterogeneity in node degrees, high clustering coefficient, and small geodesic distances. Out of the
four graph families considered in the previous sections, 
BA networks have the highest degree heterogeneity while RG
networks have the highest average clustering coefficient 
for a fixed network density (see the supplementary section and Fig.~{\color{red}\bf S2}). To gain deeper insight regarding the link between control and topology, we now
calculate the control properties of the duplex networks whose
target layers have a continuous variation in their degree heterogeneity and in their average clustering coefficient. We fix the topology of the input layer to be ER, while the target layers are constructed using a numerical scheme that generates hierarchically modular (HM) networks \cite{Chris_Nphys2020}. The network generation method has two parameters: (i) the parameter $\gamma$ that shapes the degree distribution by the mechanism of preferential attachment, and (ii) the parameter $f$ that specifies the fraction of edges in a module. The resulting networks exhibit a continuous range of degree heterogeneity and clustering coefficient (Fig.~\ref{fig:fig2}(D-E)). \\

Using these continuously varying hierarchical architectures, we calculate the optimal control energies for the target layers of duplex networks in a manner similar to that outlined in the previous section. In particular, we investigate average control energy $\bar{E}_2$ and the maximum control energy $E_{2}^{\rm{max}}$ as a function of the two generative parameters, $\gamma$ and $f$ (Figs.~\ref{fig:fig2}(F-D)). We note a similarity between the plot of maximum control energy and the plot of degree heterogeneity (Fig. \ref{fig:fig2}(D) and (G)), suggesting that the former is partially explained by the latter. The similarity between the plot of average control energy and that of degree heterogeneity is less strong; while along the $f$- axis, we note a positive correlation between the two variables, along the $\gamma$- axis this trend is inverted. We also note a negative correlation between the heatmaps of average clustering coefficient and average control energy 
(Fig.~\ref{fig:fig2}(E) and (F)). This observation is consistent with the
results we report in the previous section, where the average control energy of
the target layer was lowest for an RG network, a topology that displays the
highest average clustering among the four graph families studied (see the
supplementary section and
Fig.~{\color{red} \bf S2}). Collectively, these observations serve to support
the observed dependence between control and topology in canonical graph
families in a more continuous assessment of topological variation.

\begin{figure*}[ht]
\centering
\includegraphics[scale=0.22]{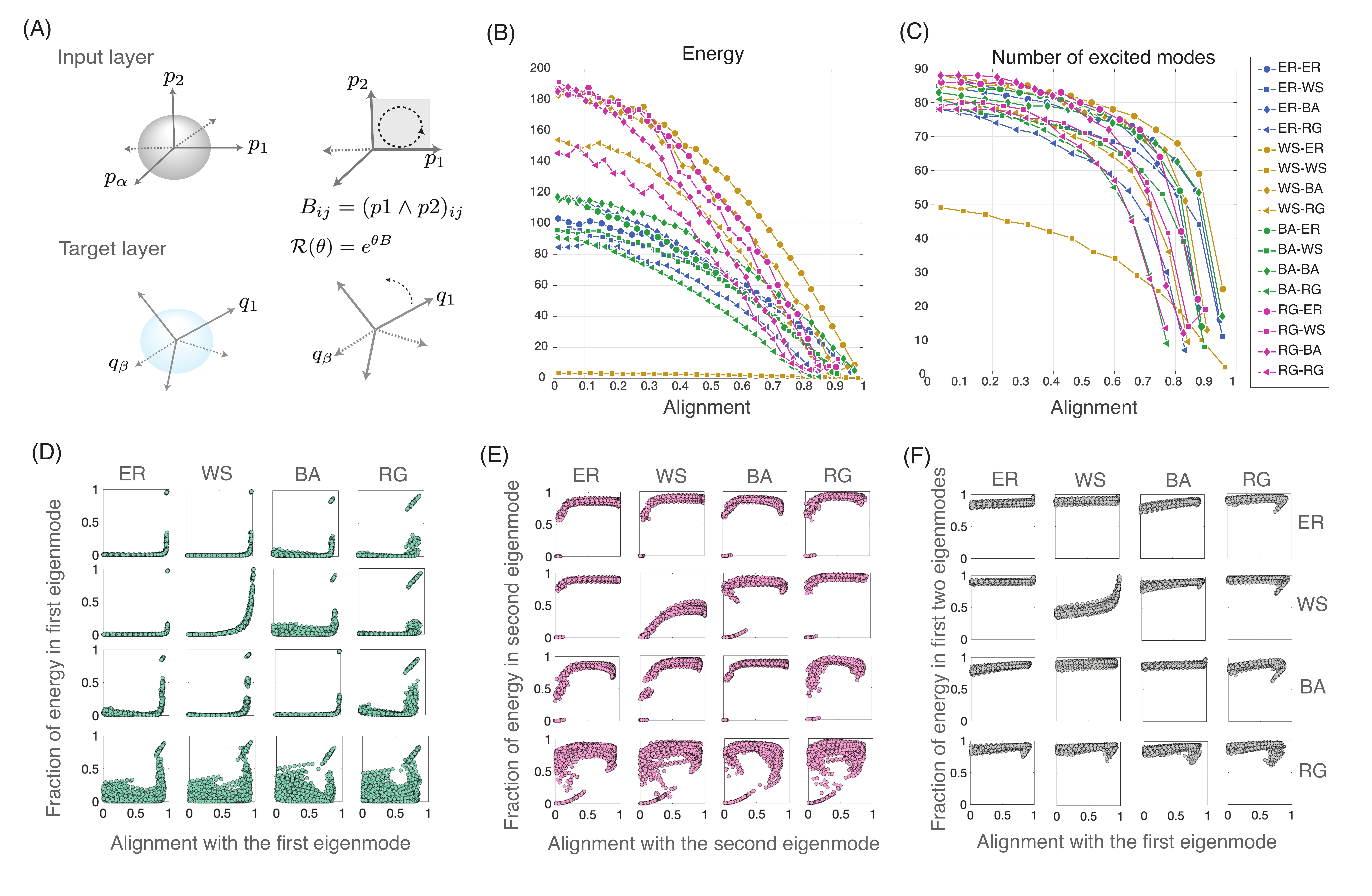}
\caption{\textbf{The role of alignment in determining the control cost.} (A) The scheme of rotation. The initial eigenspaces for each layer are shown on the left. On the right, we show how the eigenspace of the target layer is rotated relative to the input layer by defining the rotation matrix in the plane of eigenmodes corresponding to the two most positive eigenvalues of the input layer. This procedure changes the alignment of eigenmodes of the target layer only in the rotation plane, while maintaining their original alignment with the remaining (N-2)-dimensional subspace. (B) Control energy and (C) the number of excited modes to control the layer state along its dominant eigenmode as a function of alignment between the dominant eigenmodes of the two layers. The energy required to control the target layer along its dominant eigenmode is lowest when the two layers are perfectly aligned. Simultaneously, the number of eigenmodes that carry the optimal control energy is lowest when the two layers are perfectly aligned. Colors identify the topology of the input layer and markers identify the topology of the target layer. Each curve is an average of 100 simulations. (D-F) The fraction of control energy in the first (D) and the second (E) eigenmode of the input layer as a function of their alignment with the target eigenmode. (F) Total fraction of energy in the first two eigenmodes as a function of the alignment between the dominant eigenmodes of the two layers of duplex networks. 
}
\label{fig:fig3}
\end{figure*}

\subsection{The interplay between topology and alignment in determining control}
\label{subsec:mode_align}

From the analysis presented in the previous sections, we see that the alignment $\cos \theta_{\alpha \beta}$,
henceforth denoted as $C_{\alpha \beta}$, emerges as an important factor that
encodes the relative arrangement of the eigenmodes of the two layers. We now
investigate how this parameter affects the cost of controlling a specific
eigenmode in the target layer. For a clear demonstration of the role of
$C_{\alpha \beta}$, we set the control task to be the navigation of the target
layer to a final state that has a unit norm and that is aligned to the
eigenvector corresponding to the largest eigenvalue (the dominant eigenmode).
In order to explicitly control the alignment of the dominant eigenmode relative
to the eigenmodes of the input layer, we construct a rotation matrix
$\mathcal{R}(\phi)$  in the plane spanned by the eigenmodes corresponding to
the two most positive eigenvalues of the input layer (Fig.~\ref{fig:fig3}(A))
(see the supplementary section and Fig.~{\color{red} \bf S8}) \cite{ChrisDoran_GeomAlgeb}. Then, the
transformation $A_2' = \mathcal{R}(\phi)^T A_2 \mathcal{R}(\phi)$, changes the
alignment of the eigenmodes of the target layer relative to the first two
eigenmodes of the input layer. In particular, the alignment between the most
dominant eigenmodes of the two layers is predictably altered by changing the
continuous parameter $\theta$ (Fig.~\ref{fig:fig3}(A) and the supplementary
section {\color{red} \bf S8}). \\

How does the alignment of the dominant eigenmode in the target layer with that of the input layer affect the optimal cost of control? We hypothesize that when the target eigenmode is well-aligned with the dominant eigenmode of the input layer, the corresponding control energy will be small. As a corollary, the
optimal control signal will be channeled through the eigenmode of the input layer that is aligned to the dominant eigenmode of the target layer. Thus, as the dominant eigenmode of the target layer changes its alignment with the dominant eigenmode of the input layer, from a perfect alignment to an orthogonal orientation, the optimal control signal should excite an increasing number of eigenmodes in the input layer. To quantify this effect, we define $\mathcal{N}(\phi)$, which counts the number of eigenmodes $\mathbf{p}_\alpha$ of the input layer such that $\vert u_\alpha \vert^2 > 10^{-3}$. Formally, 

\begin{align}
\mathcal{N} (\phi) = \vert \{ \mathbf{p}_\alpha: \quad \vert  u_\alpha\vert^2 >
10^{-3} \} \vert .
\end{align}

\noindent We say that $\mathcal{N}(\phi)$ is the number of eigenmodes $\mathbf{p}_\alpha$ of the input layer through which the optimal control signal is `routed'. \\
 
For each duplex, we calculate the optimal energy to control the dominant eigenmode and $\mathcal{N}(\phi)$, as a function of the angle $\phi$ (see the supplementary section and Fig.~{\color{red} \bf S8}). Note that, since the rotation is defined in the plane of the first two eigenvectors of the input
layer, the alignment of the target layer relative to the remaining $N-2$ eigenvectors that form a `\emph{rotation axis}', remains unchanged. The optimal control energy as well as the number of channels that are excited in the input layer to control the dominant eigenvector of the target layer decrease as its alignment with the dominant eigenvector of the input layer decreases (Fig.~\ref{fig:fig3}(B-C)). These results imply that the alignment of a target eigenvector relative to the eigenvectors of the input layer plays an important role in determining the optimal control cost, and the eigenvectors that are excited in the process of control. \\

Note that, even as the rotation of the target eigenmode decreases the alignment with the first axis of the rotation plane, and enhances the alignment with the second axis of the rotation plane, we do not observe a decrease in energy or the number of paths. To explain this point further, as the rotating target eigenmode becomes more aligned with the second axis in the rotation plane, the optimal energy and the number of channels must decrease as the optimal energy must now be routed via the second axis. However, this effect is not observed for two reasons. First, in addition to the alignment, control energy also depends on the corresponding eigenvalue. In the analysis of the next section, we find that a more unstable eigenmode is associated with a lower energy. Thus, the second axis that corresponds to the second most unstable eigenmode has higher cost associated with it. Second, the plots of the fraction of energy carried by the first two eigenmodes show that even when the target eigenmode is aligned with the second axis, the fraction of energy routed along this axis is less than one 
(Fig.~\ref{fig:fig3}(D-F)). This effect is particularly prominent in WS and RG networks. These observations indicate that the distribution of optimal control energy, while strongly influenced by the alignment, is also affected by the topology and the eigenvalues of the corresponding eigenmodes. \\

In an LTI system with the dynamics in continuous time as considered here, eigenvalues of the adjacency matrix have the unit of a `rate'. Thus, an eigenvalue corresponding to a given eigenmode determines the time scale at which the perturbations in the corresponding eigenmodes decay or grow. The results in this section indicate that the eigenvalues, and the alignment between eigenmodes, together may determine the optimal control energy. In the next section, we attempt to understand this point further in the simplified limit of a one-mode approximation. \\

\section{Intuitions derived from a one-mode approximation}
\label{sec:onemode}

\begin{figure*}[htbp]
\centering
\includegraphics[scale=0.32]{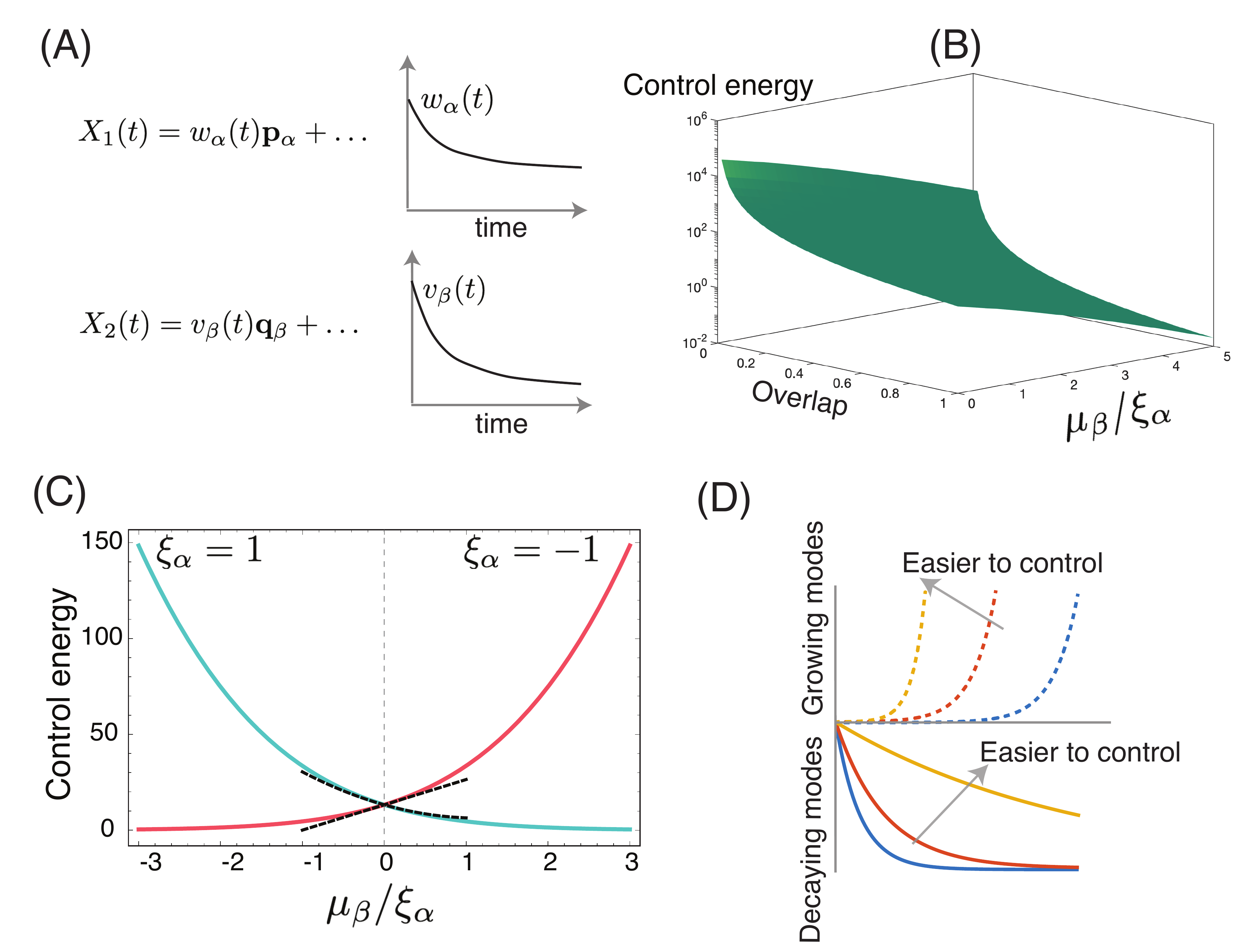}
\caption{ \textbf{Understanding structure-control relationship in single-mode limit.} (A) Control input directly couples to the input layer and can excite an arbitrary number of modes. In a one-mode approximation, we assume that the response of a layer is dominated by a single mode.  (B) Control energy required for navigating the system to the $\{0,1 \}$ state as a function both of $\mu_\beta/\xi_\alpha$ and of the
overlap between retained eigenmodes. (C) Control energy to navigate the system to a $\{0,1\}$ state as a function of $\mu_\beta/\xi_\alpha$.  (D) A faster decaying mode costs more control energy whereas a faster growing mode is easier to control. 
}
\label{fig:fig4}
\end{figure*}

To gain some insight regarding the dependence of optimal control energy on layer spectra and the interlayer alignment, we consider a simplified limit where we retain only one mode in each layer (Fig.~\ref{fig:fig4}(A)). Studying this limit can have practical applications when the phenomenon of interest happens at a specific time-scale or spatial scale. In real networks, several dynamical processes are dominated by extremal eigenvalues, such as the stability of dynamical systems on networks, synchronization, collective behaviour, and epidemic spreading \cite{Sarkar_Chaos_2018, Yang_epidemic_2003, Newman_book_2010}. Here we operationalize this simplification by revisiting Eqs.~\eqref{eq:L1-LTI} and \eqref{eq:L2-LTI-angle}. We retain the $\alpha^{\rm{th}}$ eigenmode in the input layer and the $\beta^{\rm{th}}$ eigenmode in the target layer to obtain the simplified equations: 
\begin{subequations} 
\begin{align}
\dot{w}_\alpha &= \xi_\alpha w_\alpha + U_\alpha \label{eq:1mode-L1 }\\
\dot{v}_\beta &= \mu_\beta v_\beta + \kappa \cos \theta_{\beta \alpha} w_\alpha . 
\label{eq:1mode-L2} 
\end{align}
\end{subequations}

\noindent Applying the framework of optimal control framework in this
effectively one-dimensional representation of each layer, we obtain the
solutions for the optimal state trajectory, the optimal co-state trajectory,
and the optimal control energy (see the supplementary section
{\color{red} \bf S9}). These optimal solutions depend upon the model
parameters: the eigenvalues $\xi_\alpha,\mu_\beta$, the overlap $C_{\beta
\alpha} = \cos(\theta_{\beta \alpha})$ between the retained eigenmodes, and the
strength $\kappa$ of the interlayer connections. We calculate the optimal
control energy $E^\ast $ to reach the states $\{w_{\alpha F}, v_{\beta F} \}$,
where $w \vert_{t=T} = w_{\alpha F}$ and $ v\vert_{t=T} = v_F$. \\

In order to separately calculate the control energy required to navigate the states of the two layers away from the origin, we set the final states to be $\{w_{\alpha F}, v_{\beta F} \} = \{ 1,0\}$ and $\{w_{\alpha F}, v_{\beta F} \} = \{0,1 \}$. We then consider the optimal control energy as a function of the
eigenvalues $\xi_\alpha$ and $\mu_\beta$ ( Fig.~\ref{fig:fig4}(C) and the supplementary section {\color{red} \bf S9}). We observe that a faster decay rate makes it harder to control the state of a given layer, while a faster growth rate makes it easier to control the state of that layer
along the retained eigenvector. This behavior is intuitive: the faster the decay, the harder it is to navigate the state away from the origin (Fig.~\ref{fig:fig4}(D)). \\

In the one-mode limit, we also calculate the dependence of the optimal control energy on the alignment parameter $C_{\beta \alpha}$. The dependence of control energy on $C_{\beta \alpha}$ has the form 
$E^\ast_\alpha = S_{-2} C_{\beta \alpha}^{-2} + S_{-1} C_{\beta \alpha}^{-1} + S_0$ (see the supplementary section
{\color{red} \bf S9} for detailed expressions). Here, the coefficients $S_{-2}, S_{-1}$ and $S_0$ depend on the eigenvalues corresponding to the retained eigenmodes of the two layers, 
and the final states $w_F$ and $v_F$. It is intuitive to note the dependence of the above coefficients on 
the final states $\{ w_F,v_F \}$ given by $S_{-2} \approx v_{\beta F}^2, \quad S_{-1} \approx v_{\beta F} w_{\alpha F}$ and $S_0 \approx w_{\alpha F}^2$. Note that when $v_{\beta F} \neq 0$, $E \approx C_{\beta \alpha}^{-2}$. Thus, the energy to control the state of the target layer when the control signal is
injected into the input layer diverges when $C_{\beta \alpha} \rightarrow 0$. Intuitively, it is therefore impossible to control a mode by injecting control into a mode that is orthogonal to it. Also we note that, when $v_{\beta F} =0$
but $w_{\alpha F} \neq 0$, $E^\ast_\alpha$ becomes independent of $C_{\beta \alpha}$; this independence is expected because the control is directly injected into the mode that is also being controlled. Additionally, $E^\ast_\alpha$ inevitably depends on the alignment parameter $C_{\beta \alpha}$ when $v_{\beta F} \neq 0$; that is, the control of specific eigenmodes in the
target layer is always affected by their alignment to those in the input layer. We summarize these results in Fig.~\ref{fig:fig4}(B) by plotting the optimal control energy as a function of the alignment $C_{\beta \alpha}$ and the
eigenvalues $\xi_\alpha$ and $\mu_\beta$. \\

The decrease in optimal control energy with increasing alignment is a result that we also obtained numerically in Section \ref{sec:numerics} (Fig.~\ref{fig:fig3}(B)). We note one important difference. Unlike these new observations in the one-mode limit, our numerical experiments did not show a divergence of the optimal control energy when the alignment approaches zero. This discrepancy arises from the fact that in numerically constructed networks, the eigenspaces are generally misaligned. This fact implies that even when the target eigenmode is rotated in the plane of the first two eigenvectors of the input layer, it has non-zero overlap with the remaining $N-2$ eigenvectors of the input layer, which remain unaltered by the rotation. As a result, the optimal control energy is `routed' (in the sense that is defined in Section \ref{sec:numerics}) through those eigenmodes that have non-zero overlap with the target eigenmode. \\

Finally, we ask whether the optimal control energy for the numerically
constructed duplex networks shows association with the eigenvalues of the
target eigenmodes in the second layer that are qualitatively similar to that
seen in the one-mode approximation, as depicted in Fig. \ref{fig:fig4}(C). For
all the duplex topologies and layer densities, we find an association between
the control energy of the target layer and the corresponding eigenvalues
similar to that obtained in the one-mode limit (see the supplementary section
and Fig.~{\color{red} \bf S10}). This behavior is
also generically observed for the control energy of the input layers, except in
some specific cases when input layer has a WS topology. Collectively these
findings indicate that we can gain significant understanding of the broad
relation between control and topology by performing an analytic investigation
of the simplified one-mode approximation. \\

\begin{figure*}[htp]
\centering
\includegraphics[scale=0.32]{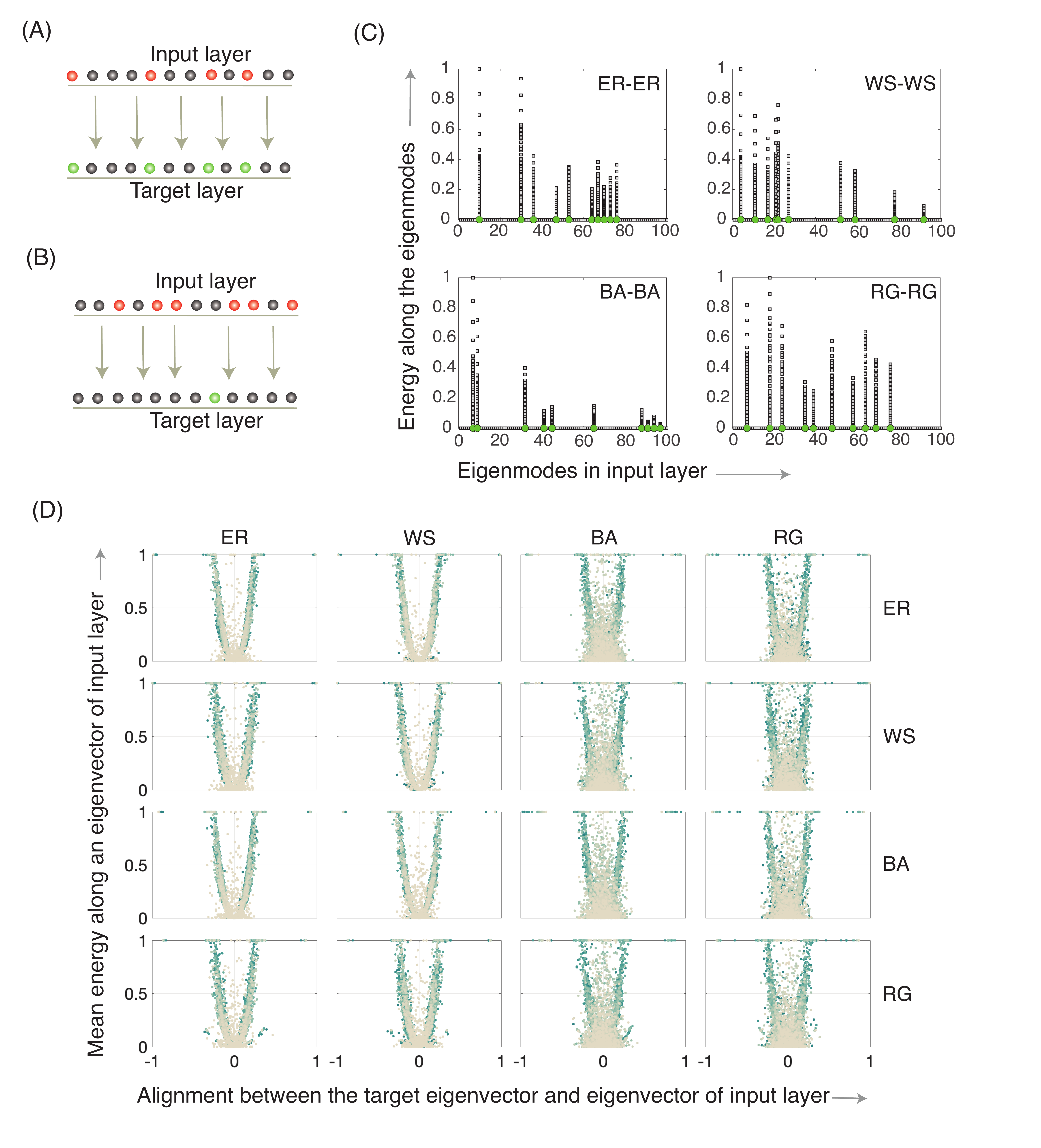}
\caption{\textbf{Distribution of control energy in the eigenmodes of the input layer.} (A) The distribution of optimal control energy into the eigenvectors of the input layer for duplex networks with identical layers. Here we construct the final state by linearly combining randomly selected target-layer
eigenmodes. We then ask whether the modes that are excited in the input layer by the optimal control signals are identical to those that constitute the target state. (B) The distribution of optimal control energy into the eigenvectors of the input layer for duplex networks with non-identical layers. Here, the final state is selected to be along the eigenvectors of the target
layer, and the time-averaged energy along each eigenvector of the input layer is calculated. (C) In the duplex network with identical layers, the eigenmodes excited in the input layer are identical to those that constitute the final state of the target layer (indicated by the green dots on $x$-axis). The instantaneous energy along the $\beta^{\rm{th}}$ eigenmode of the input layer corresponding to the $\alpha^{\rm{th}}$  eigenmode of the target layer is defined as $\vert \omega_{21}^{\alpha \beta}(t)\vert^2$. The figure shows that $\vert \omega_{21}^{\alpha \beta}(t) \vert^2$ (gray data points) is non-vanishing only for eigenmodes identical to those that constitute the final state. (D) For all of the duplex networks, the mean energy along an eigenvector of the input layer to control the state of the target layer along its eigenvectors is plotted as a function of the alignment between input eigenvectors and target eigenvectors. The mean energy is normalized by its maximum value across all of the input eigenvectors for a given target eigenvector. This normalization is performed in order to highlight the role of alignment. A strong correlation between the two quantities emerges, indicating the important role that alignment plays in control.
}
\label{fig:fig5}
\end{figure*}

\section{Path-dependence of the control energy}
\label{sec:pathdep_control}

The results obtained in the previous sections for the numerically constructed duplex networks and for the one-mode approximation indicate an important role of the alignment -- between the modes of the target layer and the modes of the input layer -- in determining the cost of control. In particular, the number of modes that are excited in the input layer increase as
the alignment between the target eigenmode and the eigenmodes of the input layer decrease. In this section, we formalize the notion of `routing' the optimal control signal further by building metrics to identify the eigenmodes that are excited in the input layer by the control signal optimized to move the state of the target layer along a given eigenvector. \\

We begin by denoting the optimal input for controlling the state of the $\alpha^{\rm{th}}$ eigenvector of the $l^{\rm{th}}$ layer by $u_l^\alpha (t)$. To measure how $u_l^\alpha$ is distributed in the eigenmodes of the $m^{\rm{th}}$ layer, we project $u_l^\alpha$ along the eigenvectors (labelled by $\beta$) of the $m^{\rm{th}}$ layer: 

\begin{align}
\mathbf{u}_{l}^\alpha(t) = \Sigma \omega_{l m }^{\alpha \beta}(t) \mathbf{V}_m^\beta , 
\label{eq:u_proj}
\end{align}

\noindent where $\omega_\beta (t)$ is the projection of the optimal input $\mathbf{u}^\alpha(t)$. For a duplex network, $l,m \in \{1,2 \}$. Then, $l=1, m=1$ corresponds to a case where the $\alpha^{\rm{th}}$ eigenmode in the input layer is the target eigenmode, and the corresponding optimal input $u^{\alpha}_1(t)$ is projected along the eigenvectors of that eigenmode. Thus,
the parameters $ \omega_{11}^\beta(t), \omega_{22}^\beta(t)$ and
$\omega_{21}^\beta(t)$ provide access to the directional information of the optimal $\mathbf{u}^\alpha(t)$. Further, the quantity $\vert \omega_{lm}^\beta(t) \vert^2$ can be identified as the energy channeled into the $\beta^{\rm{th}}$ eigenmode at time $t$. \\

The quantities $\omega^\beta_{lm} (t)$ can be utilized to identify the modes that are excited by the optimal control signal corresponding to a generic state in the state-space (i.e. not necessarily along the eigenmodes). From the insights gained in the previous sections on the role of alignment, we hypothesize that in a duplex network with identical layers, and hence perfectly aligned eigenmodes, the optimal control for a given eigenmode in the target layer will excite the identical eigenmode in the input layer. We further hypothesize that in duplex networks with non-identical layers, the fraction of the control energy along a given eigenmode of the input layer will be proportional to its alignment with the target eigenmode in the second layer. The above will be true if mode alignment is the only factor that determines the direction of optimal control. In the following we show that while the first hypothesis holds true, a test of the latter hypothesis reveals an additional (and admittedly expected) dependence on the topology of the duplex. \\

We first construct duplex networks with identical layers from all 4 graph families. We then construct a generic final state using a finite number of randomly selected eigenmodes of the target layer (Fig.~\ref{fig:fig5}(A)). We observe that the optimal control signal excites only those eigenmodes in the input layer that are identical to those forming the target state (Fig.~\ref{fig:fig5}(C)). This observation verifies the first hypothesis. To test the second hypothesis and to determine the fraction of control energy along a given eigendirection of the input layer, we calculate a time-averaged $\bar{E}_{\alpha \beta} = \vert  \omega^{\alpha \beta}_{21} (t) \vert^2$, for all $\alpha \in \{1,N \}$ (eigenmodes of the target layer) and for all $\beta \in \{1,N \}$ (eigenmodes of the input layer). The time-averaged $\bar{E}_{\alpha \beta}$ then represents the mean energy along the $\alpha$ eigenvector corresponding to the control of the target layer along the $\beta^{\rm{th}}$ eigenvector. We assess the correlation between $\bar{E}_{\alpha \beta}$ and the alignment of the target eigenmode with the
eigenmodes of the input layer for all the constructed duplex networks and all densities (Fig. \ref{fig:fig5}(C)). We find that the alignment indeed emerges as a strong factor in determining the fraction of control energy along an eigenvector of the input layer. This fact is evidenced by the presence of a strong trend of 
$\bar{E}_{\alpha \beta} \propto f(\vert C_{\alpha \beta}\vert)$ where $f$ is an increasing function of the alignment $ \vert C_{\alpha \beta} \vert$. However, deviations from this behavior appear especially for higher densities (Fig. \ref{fig:fig5}(C)), and when the target layer has the topology of a BA or an RG network.

\section{Conclusion and Future Directions} 
\label{sec:conclusion&future}

In our investigation, we have sought to better understand the relationship between structure and the cost of control, in a principled and theoretically motivated way. Our efforts, which combine analytical results and numerical experiments, allow us to build a generic framework to extract the dependence of multiplex control on layered architecture. We find that the response rates as characterized by the eigenspectra of each layer, and the alignment between the eigenmodes of the input and target layers emerge as key parameters that determine the cost of optimal control as well as the distribution of optimal control energy in different channels. While results from a one-mode approximation provide some useful insight regarding the correlation between eigenvalues and control cost of individual eigenmodes, the global characteristics of overall network topology (such as density, degree distribution, and clustering coefficient) determine the specific manner in which optimal control energy is distributed in the eigenmodes of the two layers. The effect of the global topology is particularly prominent in RG and WS networks, both of which show high values of average clustering coefficient. Our work indicates that a high value of average clustering coefficient is associated with lower average control energy while a high value of degree heterogeneity is associated with the maximum control energy. 
In addition, we also studied the distribution of optimal control energy into the eigenmodes of the input layer in interlayer control. Our approach can be utilized in identifying and predicting the paths of optimal control in specific applications. \\

In summary, we investigate the dependence of control properties of layers in duplex networks as they are dictated by layer spectra and the alignment between different eigenmodes of the two layers. We establish the key role of mode alignment in controlling the state of the target layer along specific eigendirections and further highlight the interplay between alignment and layer topologies in determining the cost and routing of optimal control energies. Our approach of 
determining the optimal energies for state transitions in a \emph{layer-diagonalised} representation can be easily generalized to multiplex networks with more than two layers, and to more generic interlayer connections. The specific manner in which a more generic pattern of interlayer connections interplays with the relative alignment between the input and the target layers is an exciting direction of future research in the theory of multiplex control. Our results will serve as a platform for future work in narrowing the space of feasible intervention strategies in applications where cross-layer control is required. Indeed, recent multilayer representations of networks in biology have already indicated that different connectivity layers can be misaligned \cite{Becker_SciRep2018}, impacting the system's potential response to perturbations. Using our framework, future work can explore the consequences of such layered architectures on designing optimal interlayer control strategies.

\section{Citation Diversity Statement}

Recent work in several fields has identified a bias in citation practices such
that papers from women and other minorities are under-cited relative to the
number of such papers in the field \citep{Dworkin2020.01.03.894378,
maliniak2013gender, caplar2017quantitative,
chakravartty2018communicationsowhite,
YannikThiemKrisF.SealeyAmyE.FerrerAdrielM.Trott2018, dion2018gendered}. Here we
sought to proactively consider choosing references that reflect the diversity
of the field in thought, gender, race, ethnicity, and other factors. We
obtained the predicted gender of the first and last author of each reference by
using databases that store the probability of a first name being carried by a
woman \citep{Dworkin2020.01.03.894378}. By this measure (and excluding
self-citations to the first and last authors of our current paper), our
references contain $ 60 \%$ man/man, $20 \%$ man/woman, $4.4 \%$ woman/man, $15.6 \%$
woman/woman, and $0 \%$ unknown categorization. Open-source code for this
estimation process is publicly available \citep{github_dale}. This method is
limited in that a) names, pronouns, and social media profiles used to construct
the databases may not, in every case, be indicative of gender identity and b)
it cannot account for intersex, non-binary, or transgender people. We look
forward to future work that could help us to better understand how to support
equitable practices in science.

\acknowledgments
This work was primarily supported by the Schwartz Foundation, the Army Research
Office (Falk-W911NF-18-1-0244), and the Paul G. Allen Family Foundation. We
would also like to acknowledge additional support from the John D. and
Catherine T. MacArthur Foundation, the Alfred P. Sloan Foundation, the ISI
Foundation, the Army Research Office (Grafton-W911NF-16-1-0474,
DCIST-W911NF-17-2-0181), the National Institute of Mental Health
(2-R01-DC-009209-11, R01-MH112847, R01-MH107235, R21-M MH-106799), and the
National Science Foundation (NSF PHY-1554488, BCS-1631550, and IIS-1926757).
The content is solely the responsibility of the authors and does not
necessarily represent the official views of any of the funding agencies. 

\bibliography{Refs_multicontrol_theory}
\bibliographystyle{ieeetr}

\end{document}


\title{Structural underpinnings of control in multiplex networks: Supplemental Material}

\date{\today}
\maketitle

\tableofcontents

\newpage
\section
{Optimal Control Theory for Duplex Networks: Exact Solution}
\label{Ssec1:exact_sol}

In this section, we derive the exact expressions for the state variables and control signals for the control problem of a duplex network where control inputs are injected only into one layer. We begin by rewriting the minimization problem in Eq.(6) 
of the main text: 

\begin{align}
\underset{U(t)}{{\rm{minimize}}} \quad E, \nonumber\\
\text{subject to }  \dot{w}_\alpha &= \xi_\alpha w_\alpha + \mathbf{b}_\alpha
\cdot U \nonumber \\
\dot{v}_\alpha &= \mu_\alpha v_\alpha + \kappa
\underset{j}{\Sigma} w_\beta \cos \theta_{\alpha \beta} .
\label{Seq:optim_statement} 
\end{align}

\noi
Here, the energy $E$ is written in terms of the control input as $E = \int_0^{t_F} \mathbf{u(t)}^T \mathbf{u(t)} dt$, and the quantity $t_F$ denotes time horizon of the control problem.  In addition, the boundary conditions are specified as $w_\alpha(0) = 0, w_\alpha(t_F) = w_{\alpha F}, v_\alpha = 0, $ and $ v_\alpha (t_F) = v_{\alpha F}$ for all $\alpha$. We solve the above minimization problem in the framework of optimal control theory \cite{lewis_optimal_control_2012}. The dynamic constraints can be included in the definition of an augmented cost function $J$ using Lagrange multipliers, written as:

\begin{align}
J  &= \int_0^T (\U^T \U + \boldsymbol{\lambda}(t)^T (D_1 \mathbf{w} + \U - \dot{\mathbf{w}}) +  \boldsymbol{\sigma}(t)^T (D_2 \mathbf{v} + \mathbf{C} \mathbf{v} - \dot{\mathbf{v}}(t)) ) dt \\
&= \int_0^T (H(\w,\mathbf{v},\mathbf{u}) - \boldsymbol{\lambda}(t)^T \dot{\mathbf{w}}(t) - \boldsymbol{\sigma}(t)^T \dot{\mathbf{v}}(t) ) . 
\end{align}

\noi
Here, we have used two sets of Lagrange multipliers, also called co-state variables, $\boldsymbol{\lambda}(t) \in \RR^N$ and $\boldsymbol{\sigma}(t) \in \RR^N$, which introduce a cost on the deviation of the system's dynamics from the state equations. Additionally, we have written the state equations in their vector-form: $ \dot{\mathbf{w}} = D_1 \mathbf{w}  + \U $, and $\dot{\mathbf{v}} = D_2 \mathbf{v}+ \mathbf{C} \mathbf{w}$. In writing $J$, we have also used the form of the input matrix
$\mathbf{B}$. Finally $\mathbf{C}_{ij} =  p_i \mathbf{K} q_j$, where $\mathbf{K}$ denotes the interlayer connectivity and is written as $\mathbf{K} = \kappa \mathbf{I}$. Thus,  $\mathbf{C}_{ij}= \kappa \cos \theta_{ij}$. \\

Identifying a part of $J$ as the Hamiltonian, $H$, we can further write the full cost function $J$ as: 

\begin{align}
J&= \int_0^T (H(\w,\mathbf{v},\mathbf{u}) - \boldsymbol{\lambda}(t)^T \dot{\mathbf{w}}(t) - \boldsymbol{\sigma}(t)^T \dot{\mathbf{v}}(t) ).
\label{eq::J_augmented}
\end{align}

\noi
Here, 
\begin{align}
H &= \U^T \U + \boldsymbol{\lambda}(t)^T (D_1 \mathbf{w} + \U ) +  \boldsymbol{\sigma}(t)^T (D_2 \mathbf{v} + \mathbf{C} \mathbf{v}) .
\label{eq:hamiltonian}
\end{align}

\noindent The necessary conditions for the solution of the above minimization problem can then be obtained using Pontryagin's maximum principle \cite{lewis_optimal_control_2012}, which requires the variation of $J$ with respect to increments in state variables, co-state variables and independent variables to vanish. As a result, we obtain the following for the optimal co-state variables:

\begin{align}
\dot{\lambda}^{\ast T} &= - \f{\p H}{ \p \w} = - \left(\boldsymbol{\lambda}^{\ast T} D_1 
+ \boldsymbol{\sigma}^{\ast T} \mathbf{C} \right)
\label{lam_dyn}\\
\dot{\boldsymbol{\sigma}}^{\ast T} &= - \boldsymbol{\sigma}^{\ast T} D_2 .
\label{sig_dyn}
\end{align}

\noindent Here the superscript $\ast$ denotes the optimal solution. Next, the optimality condition gives a relationship between the optimal input $\mathbf{u}^\ast$ and the co-state
variable $\boldsymbol{\lambda}^\ast$:

\begin{align}
0 &=\f{\p H}{ \p U} \bigg\vert_{U^\ast} \nonumber \\
  &= 2 \U^\ast +  \lambda^\ast \nonumber \\
\U^\ast &= -  \boldsymbol{\lambda}^\ast/2 .
\label{eq:ustar}
\end{align}

The Eqs. \eqref{lam_dyn} and \eqref{sig_dyn}, along with the state equations, together form $4N$ equations for $4N$ unknowns. With the $4N$ boundary conditions given by the initial and final conditions for $\mathbf{w}$ and $\mathbf{v}$, the system of equations is determined, and can be solved exactly. Starting with the equation \eqref{sig_dyn}, we solve and substitute for $\boldsymbol{\lambda}$, $\mathbf{w}$, and $\mathbf{v}$ successively. These solutions are given as: 

\begin{align}
\boldsymbol{\sigma}^\ast(t) &= e^{-D_2 t} \boldsymbol{\sigma}_0, .
\label{eq:sigopt_sol}
\end{align}

\noi
Substituting this expression in the equation for $\lambda$, we obtain 

\begin{align}
\dot{\boldsymbol{\lambda}}^\ast(t) &= - D_1 \boldsymbol{\lambda}^\ast
- \mathbf{C}^T e^{-D_2 t} \boldsymbol{\sigma}_0.
\label{eq:lam_eq}
\end{align}

\noi 
Writing component-wise, we obtain

\begin{align}
\dot{\lambda}^\ast_i(t) &= -\xi_i \lambda_i^\ast - \Sigma_j C_{ji} e^{-\mu_j t} \sigma_{0j} .
\label{eq:lamcomp_eq}
\end{align}

\noindent Integrating the above expression, we obtain the following for $\lambda_i$

\begin{align}
\lambda^\ast_i(t) &= e^{- \xi_i t} \l \lambda_{0i} + \Sigma_j \sigma_{0 j} C_{ji} \f {e^{- (\mu_j - \xi_i) t} -1 }{\mu_j - \xi_i } \r \nn\\
&= e^{- \xi_i t}\lambda_{0i} + \Sigma_j \sigma_{0j} C_{ji} \f{e^{-\mu_j t} - e^{-\xi_i t}}{ \mu_j -\xi_i} .
\label{eq:lamopt_sol}
\end{align}

\noindent Substituting the solution for $\boldsymbol{\lambda}^\ast$ in the equation for the 
state variables for first layer, we obtain

\begin{align}
\dot{w}_i(t) =  \xi_i w_i - \f{\lambda_i^\ast (t)}{2} . 
\label{eq:w_eq}
\end{align}

Then, the solution for the optimum trajectory for the state of the first layer is given by 
\begin{align}
w^\ast_i(t) &=  e^{\xi_i t} w_{0i} -  \f{e^{\xi_i t}}{2} \int_0^t \lambda_i(\tau) e^{-\xi_i \tau} d \tau \nn\\
&= e^{\xi_i t} w_{0i} - \f{1}{2} e^{\xi_i t} \int_0^t  \l  e^{-\xi_i \tau} \lambda_{0i} + \Sigma_j \sigma_{0j} C_{ji} \frac{ e^{-\mu_j t} - e^{-\xi_i t}}{ \mu_j - \xi_i} \r e^{-\xi_i \tau} \nn\\
&= e^{\xi_i t} w_{0i} + \f{\lambda_{0i}}{4 \xi_i} (e^{- \xi_i t} -e^{\xi_i t})  + 
\Sigma_j \f{\sigma_{0j} C_{ji} e^{\xi_i t} }{ 4(\mu_j -\xi_i)} \left[ \f{e^{-(\xi_i +\mu_j) t} -1 }{ \xi_i + \mu_j }  - \f{e^{-2 \xi_i t} -1}{ 2 \xi_i }\right] \nn\\
&= e^{\xi_i t} w_{0i} - \f{\sinh(\xi_i t)}{ 2 \xi_i} \lambda_{0i} 
+ \Sigma_j \f{ C_{ji}}{ 2 (\mu_j - \xi_i )} \left[ \f{e^{-\mu_j t} - e^{\xi_i t} }{ \xi_i + \mu_j }  + \f{\sinh(\xi_i t)}{ \xi_i}\right] \sigma_{0j} .
\label{eq:wopt_sol}
\end{align}

\noindent Using the expressions of $w^\ast_i(t)$ in the state equation for the second layer, we find

\begin{align}
\dot{v}_i(t) = \mu_i v_{i} + \Sigma_j C_{ij} w_j(t) .
\label{eq:v_eq} 
\end{align}

\noindent Finally, the solution for the optimal state trajectory of the second layer is given by 

\begin{align}
v^\ast_i(t) &= e^{\mu_i t} v_{i0} + \Sigma_j C_{ij} e^{\mu_i t} \int_0^t w_j(\tau) e^{-\mu_i \tau} d \tau  \nn\\
&= e^{\mu_i t} v_{i0} 
 + \Sigma_j C_{ij} \f{ e^{\xi_j t} - e^{\mu_i t}}{ \xi_j - \mu_i} w_{0j} 
 - \Sigma_j \f{C_{ij}}{ 4 \xi_j} \left[  \f{e^{-\xi_j t} - e^{\mu_i \tau}}{ \xi_j + \mu_i} + \f{e^{\xi_j t} - e^{\mu_i t}}{ \xi_j - \mu_i } \right]  \lambda_{0j} \nn\\
& 
- \Sigma_{j,k} \f{C_{ij} C_{kj}}{ 2 (\mu_k^2 - \xi_j^2)} \left[  \f{e^{-\mu_k t} - e^{\mu_i t}}{ \mu_k + \mu_i}  + \f{e^{\xi_j t} - e^{\mu_i t}}{ \xi_j - \mu_i}  \right] \sigma_{0k} 
+ \Sigma_{j,k} \f{C_{ij} C_{kj}}{ 4 \xi_j (\mu_k - \xi_j)} \left[  \f{e^{\xi_j t} - e^{\mu_i t}}{ \xi_j - \mu_i}  + \f{e^{-\xi_j t} - e^{\mu_i t}}{ \xi_j + \mu_i}  \right] \sigma_{0k} .
\label{eq:vopt_sol}
\end{align}

The above solutions for the co-state and state variables in Eqs. \eqref{eq:sigopt_sol}, \eqref{eq:lamopt_sol}, \eqref{eq:wopt_sol}, 
and \eqref{eq:vopt_sol}can be further expressed in a compact matrix form. Making use of the initial state of the state variables, the solutions can be written as: 

\begin{subequations}
\begin{align}
\sigma^\ast(t) & = M^\sigma (t) \boldsymbol{\sigma}_0
\label{eq:sol_mat1} \\
\lambda^\ast(t) &= M^\lambda(t) \boldsymbol{\lambda}_0 
\label{eq:sol_mat2}\\
w^\ast(t) &= -G(t) \lambda_0 + H(t) \boldsymbol{\sigma}_0
\label{eq:sol_mat3} \\
v^\ast(t) &= - G'(t) \lambda_0 + H'(t) \boldsymbol{\sigma}_0 ,
\label{eq:sol_mat4} 
\end{align}
\end{subequations}

\noindent where the coefficient matrices are given by 

\begin{subequations}
\begin{align}
M^\sigma_{ij}(t) &= \exp(-\mu_i t) \delta_{ij} 
\label{eq:coeff_M1}\\
M^\lambda_{ij}(t) &= \exp(-\lambda_i t) \delta_{ij} 
\label{eq:coeff_M2}\\
G_{ij}(t) &= \f{\sinh(\xi_i t)}{ 2 \xi_i} \delta_{ij} 
\label{eq:coeff_M3}\\
H_{ij}(t) &= \f{C_{ji}}{ 2 (\mu_j - \xi_i)} \left[ \f{e^{-\mu_j t} - e^{\xi_i
t}}{ \xi_i + \mu_j} + \f{ \sinh{(\xi_i t)}}{ \xi_i}\right] 
\label{eq:coeff_M4}\\
G'_{ij}(t) & =   \f{C_{ij}}{ 4 \xi_j} \left[  \f{e^{-\xi_j t} - e^{\mu_i \tau}}{ \xi_j + \mu_i} + \f{e^{\xi_j t} - e^{\mu_i t}}{ \xi_j - \mu_i } \right]  \lambda_{0j} 
\label{eq:coeff_M5}\\
H'_{ij}(t) &= \Sigma_{k} \f{C_{ik} C_{jk}}{ 2 (\mu_j^2 - \xi_k^2)} \left[  \f{e^{-\mu_j t} - e^{\mu_i t}}{ \mu_j + \mu_i}  + \f{e^{\xi_k t} - e^{\mu_i t}}{ \xi_k - \mu_i}  \right] 
+ \Sigma_{k} \f{C_{ik} C_{jk}}{ 4 \xi_k (\mu_j - \xi_k)} \left[  \f{e^{\xi_k t} - e^{\mu_i t}}{ \xi_k - \mu_i}  + \f{e^{-\xi_k t} - e^{\mu_i t}}{ \xi_k + \mu_i}  \right]  .
\label{eq:coeff_M6}
\end{align}
\end{subequations}

At this stage, the initial conditions for the co-state variables remain unknowns and can be determined using the boundary conditions for the state variables at the terminal time $t_F$. This fact yields the following expression for $\lambda_0$ and $\sigma_0$:

\begin{subequations}
\begin{align}
\boldsymbol{\lambda}_0 &= - G(t_F)^{-1}
( \mathbf{w}_F - H(t_F) \boldsymbol{\sigma}_0 ) 
\label{eq:lam_init_eq} \\
\boldsymbol{\sigma}_0 &= H(t_F)^{-1} \l v_F + G'(t_F) \boldsymbol{\lambda}_0 \r  .
\label{eq:sig_init_eq}
\end{align}
\end{subequations}

\noindent Solving these, we obtain the initial values of the co-state variables: 

\begin{subequations}
\begin{align}
\boldsymbol{\lambda}_0 &= \frac{\mathbf{v}_F - H'(t_F) H(t_F)^{-1} \mathbf{w}_F}
{ - G'(t_F) + H'(t_F) H(t_F)^{-1} G(t_F)} 
\label{eq:lam_init_sol} \\
\sigma_0 &= H^{-1} (\mathbf{w}_F + G(T) \boldsymbol{\lambda}_0) .
\label{eq:sig_init_sol}
\end{align}
\end{subequations}

\begin{figure}[h]
    \centering
    \includegraphics[scale=0.4]{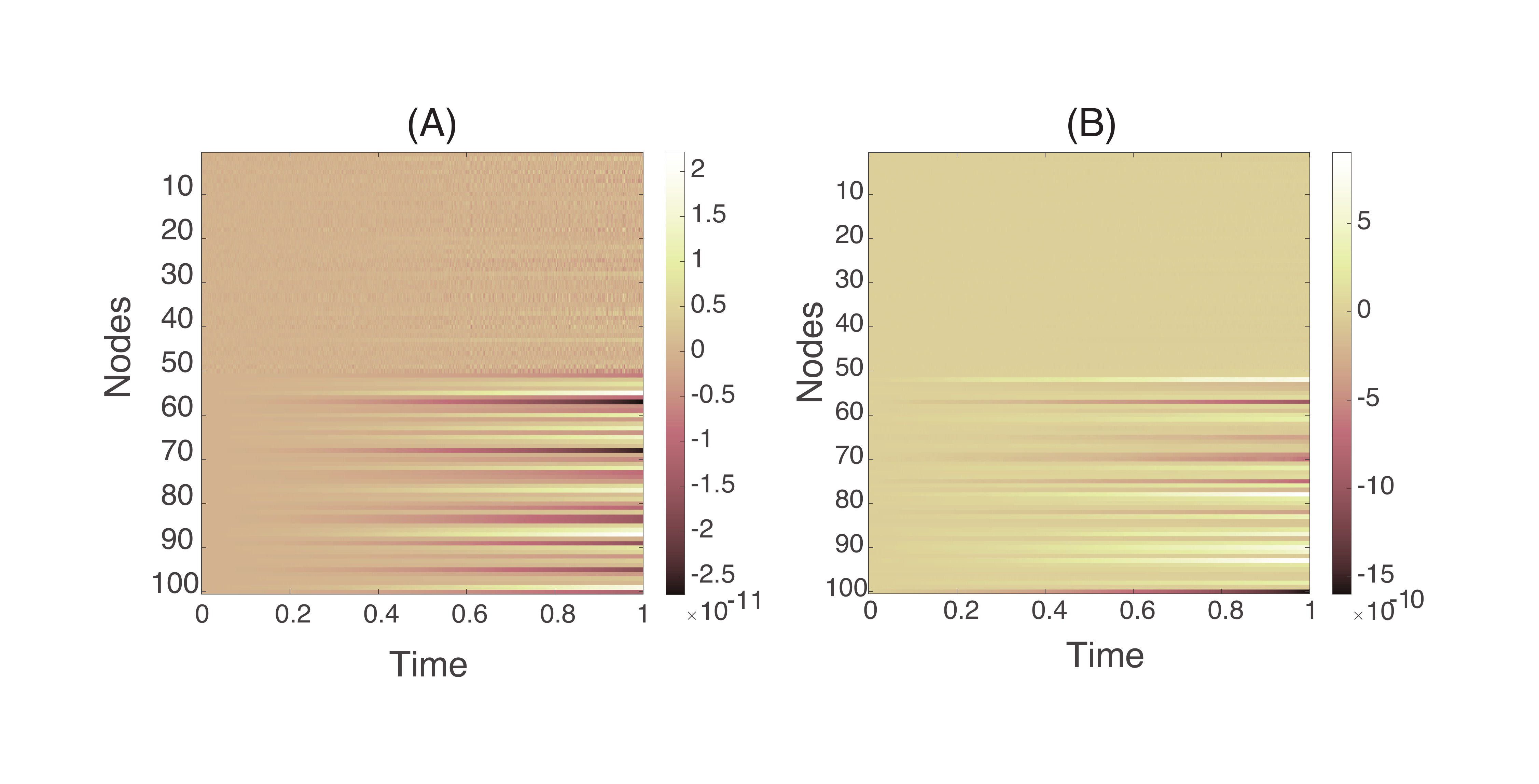}
    \caption{\textbf{Comparison of analytical and numerical solutions for state and costate variables.} A heatmap of differences between the numerical calculations and the analytical solutions for (A) state variables (Eqs. \eqref{eq:sol_mat3} and \eqref{eq:sol_mat4}) and
    (B) co-state variables (Eqs. \eqref{eq:sol_mat1} and \eqref{eq:sol_mat2}).}
    \label{fig:sol_compare}
\end{figure}

\noi
\Crefrange{eq:sol_mat1}{eq:sol_mat4}, with the coefficient matrices
(\Crefrange{eq:coeff_M1}{eq:coeff_M6}) 
and the initial values of the co-state variables (\Crefrange{eq:lam_init_sol}
{eq:sig_init_sol}) complete the solution for the optimum state and co-state trajectories for duplex networks. Finally, the optimal control input can be determined from the expression of 
$\boldsymbol{\lambda}^\ast$ using Eq. \ref{eq:ustar}. We verify that the above analytical solutions match the direct numerical calculations. We find that the initial conditions for the co-state variables matched within $\sim 10^{-7}$, while the analytical solutions for the state and co-state variables matched within $\sim 10^{-10}$ (Fig. \ref{fig:sol_compare}).

\section
{Topological properties of networks}
\label{Ssec2:topo_prop_of_nets}

\begin{figure}[h]
    \centering
    \includegraphics[scale=0.35]{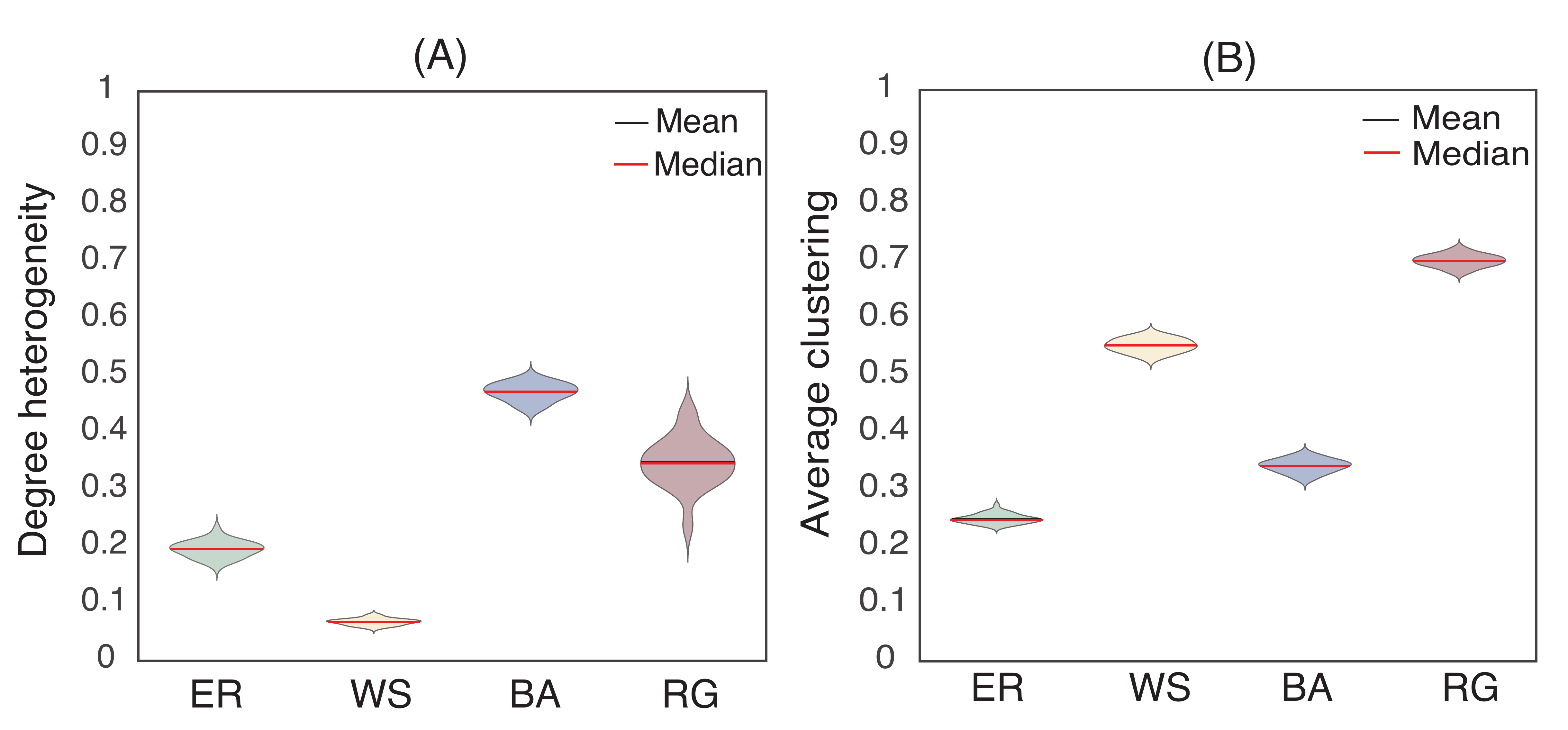}
    \caption{\textbf{Comparison of topological properties of networks from the four graph families.} Here we show violin plots of the (a) degree heterogeneity and (b) average clustering clustering coefficient for 40 simulations of each network.}
    \label{fig:fig2_topo4}
\end{figure}

In this section, we examine two metrics of heterogeneity -- degree heterogeneity and average clustering coefficient -- for networks drawn from the four graph families studied in detail in the main paper: Erd\H{o}s-R\'{e}nyi (ER),  Watts-Strogatz(WS),  Barab\'{a}si-Albert (BA), and Random Geometric (RG). We calculate the degree heterogeneity and the average clustering for 40 simulations of each graph model for a fixed network density $\rho = 0.25$. We find that the degree heterogeneity is highest for the BA network model, whereas average clustering coefficient is highest for the RG network model (Fig. \ref{fig:fig2_topo4}).

\section{Generating networks of a given density} 
\label{Ssec3:gennets_dens}
\begin{figure}[h]
    \centering
    \includegraphics[scale=0.55]{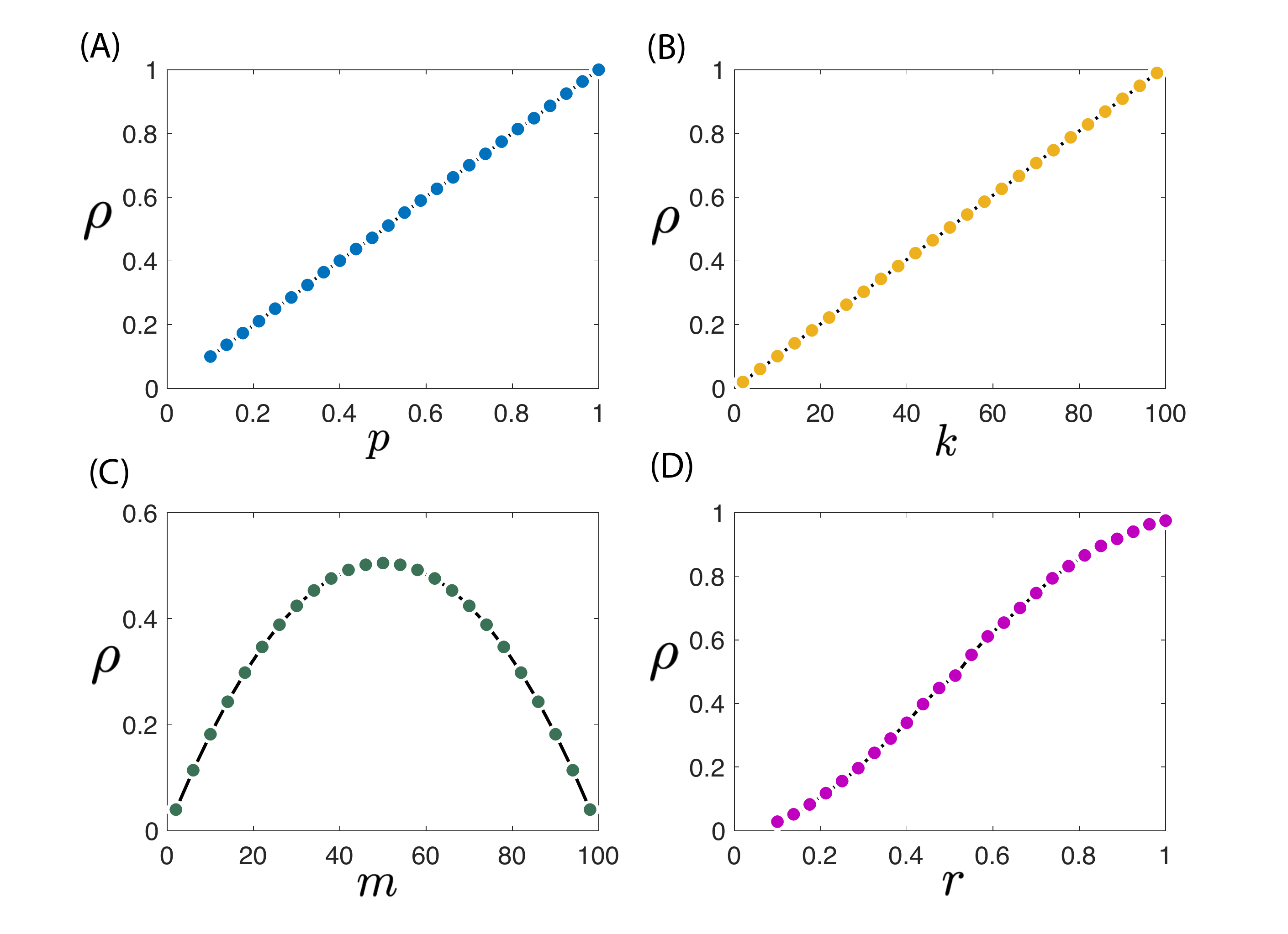}
    \caption{\textbf{Network density $\rho$ as a function of the generating parameters of graph families.} (A) The density of an ER network as a function of the probability of edge creation $p$. (B) The density of a WS network as a function of the number of nearest neighbors fixed by the parameter $k$. (C) The density of a BA network as a function of the number of edges $m$ that a new node forms with the existing nodes based on preferential attachment. Note the non-monotonic dependence of density as a function of $m$ leading to loops in the plots of average and maximum control energy when the target layer has the topology of a BA network (Fig.~\ref{fig:fig2}(B) and (C)). (D) The density of an RG network as a function of the radius parameter $r$.}
    \label{fig:fig3_NetDens}
\end{figure}

To construct the duplex networks studied in Section {\color{red} \bf III} of the main text, we generate networks from the four graph families:  Erd\H{o}s-R\'{e}nyi (ER),  Watts-Strogatz(WS),  Barab\'{a}si-Albert (BA), and Random Geometric (RG). We use
the standard NetworkX algorithms to generate these networks, and study their network density $\rho$ as a function of their respective generating parameters (Fig. \ref{fig:fig3_NetDens}).
We define the density $\rho$ as $\rho= e/N(N-1)$ where $e$ is the number of edges in the network and $N$ is the number of nodes.We observe a non-monotonic dependence of density  on the generating parameter $m$ for BA networks; BA networks display the highest density when $m = N/2$, where $N$ is the size of the network. This behavior is intuitive: for $m < N/2$, the number of edges a new node forms with the existing nodes is small, whereas for $m>N/2$, even though a new node has the tendency to form a large number of edges, it cannot, as the number of existing nodes is less than $m$.

\section{The distribution of optimal energy}
\label{Ssec4:optimE_dist}

\begin{figure}[h]
    \centering
    \includegraphics[scale=0.33]{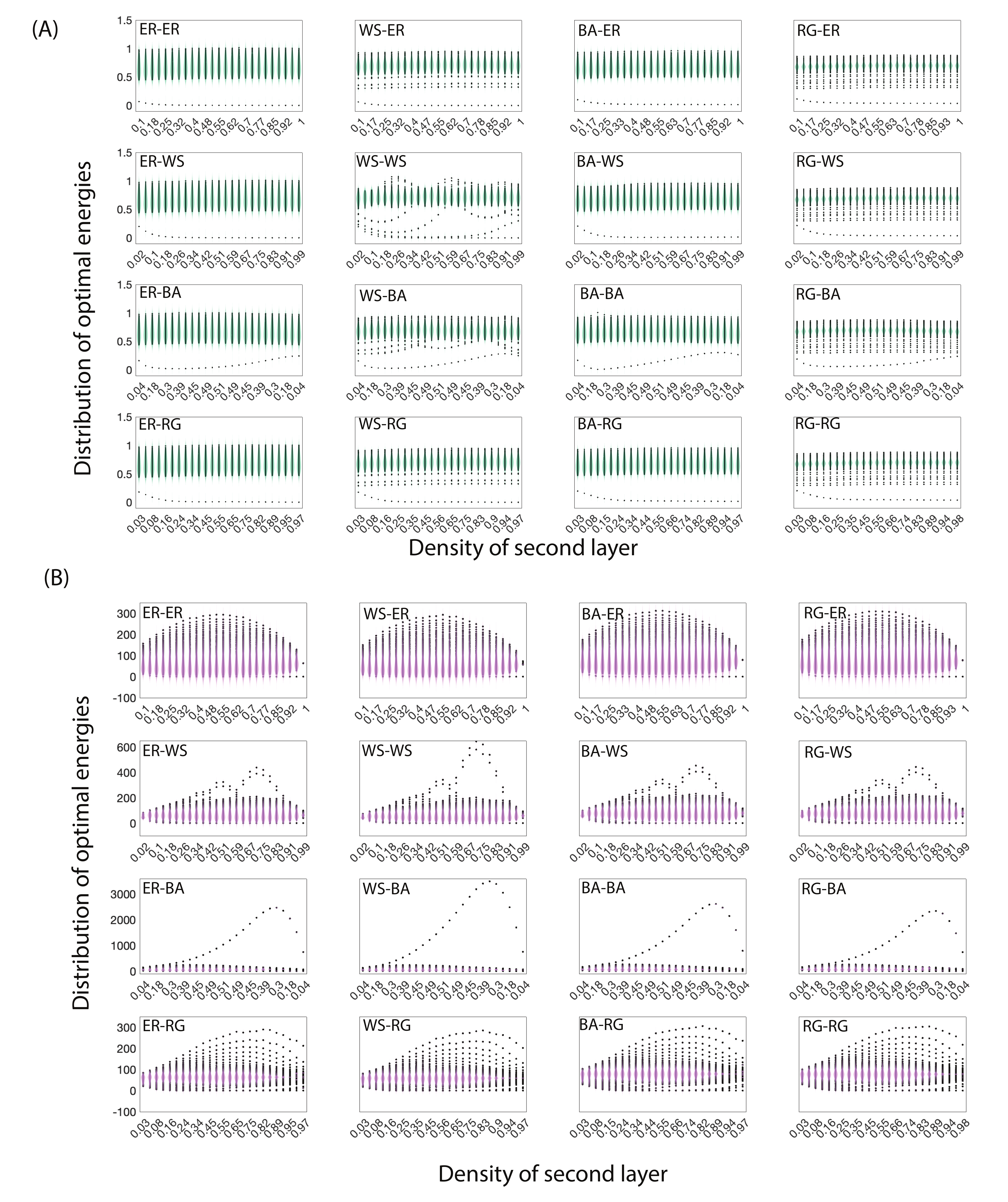}
    \caption{\textbf{Spread of optimal control energy values.} The 
    optimal control energies to move (A) the state of the first layer, and (B) the state of the second layer, along all of their eigendirections as a function of network density. The spread of energy values is displayed by the spread of data points along y-axis for each layer, overlaid with violin plots. Each curve is an average of 20 realizations.}
    \label{fig:fig4_EDist}
\end{figure}

In this section, we examine the distribution of optimal control energies for
all of the duplexes considered in Section {\color{red} \bf III A} of the main
text. For each duplex network, we fix the density of the input layer and change
the density of the second layer. We calculate the optimal control energy for
moving the state of each layer along its eigendirections by a unit amount (Fig.
\ref{fig:fig4_EDist}). We note that for the first layer, the minimum optimal
energy separates from the bulk for all of the duplexes except for WS-WS and
WS-RG combinations, where more than one energy value separates from the bulk.
This behavior indicates that for WS and RG networks, the first few
eigendirections cost much less energy to control than the bulk of
eigendirections. This observation is consistent with the shape of the
underlying eigen-spectra of the WS and RG networks, where more than one
eigenvalue separates from the bulk \cite{Sarkar_Chaos_2018}.

\section{The effect of interlayer connection strength on target-layer control}
\label{Ssec5:kappa_effect}
\begin{figure}[h]
    \centering
    \includegraphics[scale=0.4]{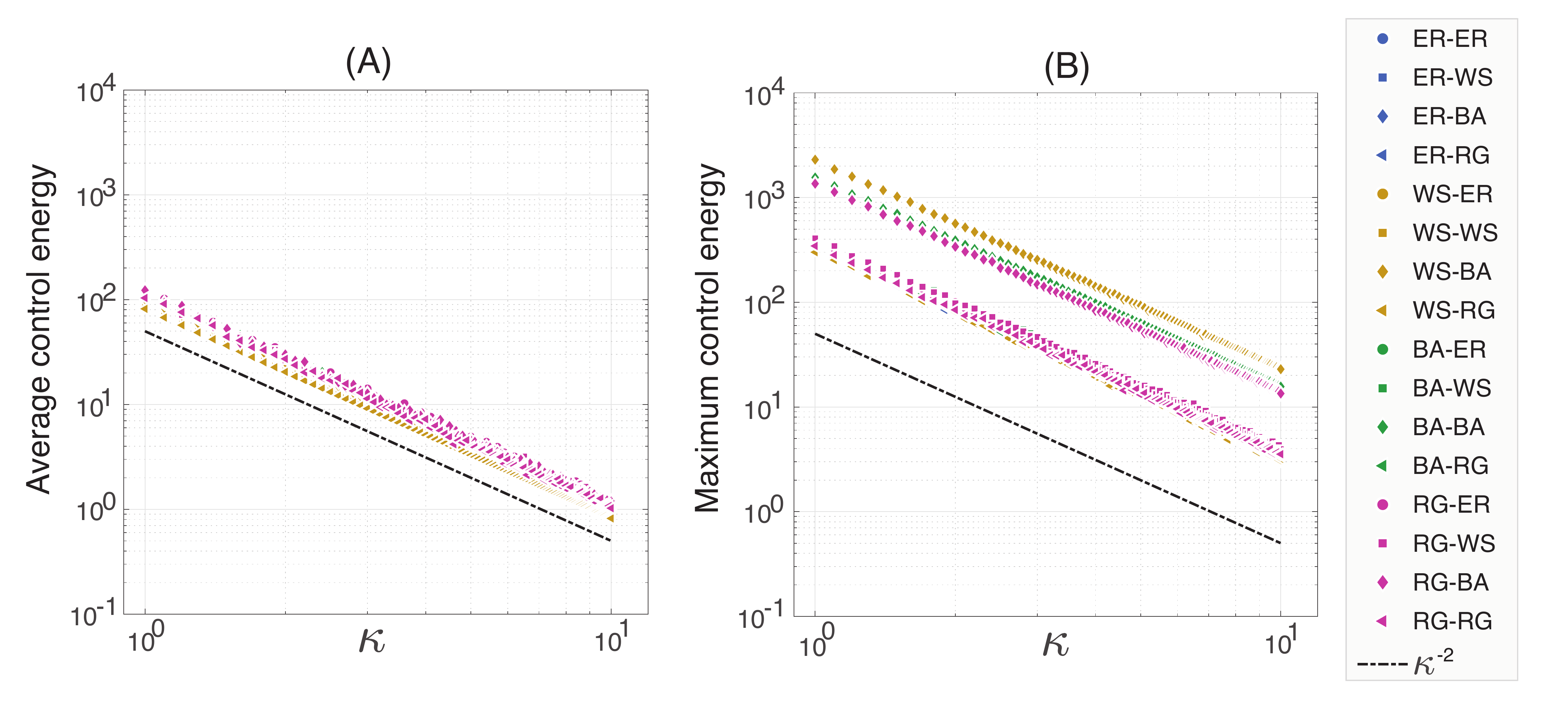}
    \caption{\textbf{The effect of the interlayer connection strength on the target-layer control.} (A) Average control energy and (B) maximum control energy as a function of $kappa$, the interlayer connection strength. The black dotted line in both panels is a plot of $\kappa^{-2}$, a trend predicted by the analytical results in the single-mode limit.}
    \label{fig:fig5_roleofkappa}
\end{figure}

As outlined in the main text, in a continuous-time model of linear dynamics, the strength of the interlayer connection, $\kappa$, is an important parameter that regulates the rate of the propagation of a given perturbation from the input layer to the target layer. We therefore examine the average and maximum control energy as a function of $\kappa$ for all of the different topologies of duplex networks studied in this work (Fig. \ref{fig:fig5_roleofkappa}). We find that an increasing strength of the interlayer connections decreases the average and maximum control energies.

\section{The effect of input-layer density}
\label{Ssec6:input_layer_dens_effect}
\begin{figure}
    \centering
    \includegraphics[scale=0.35]{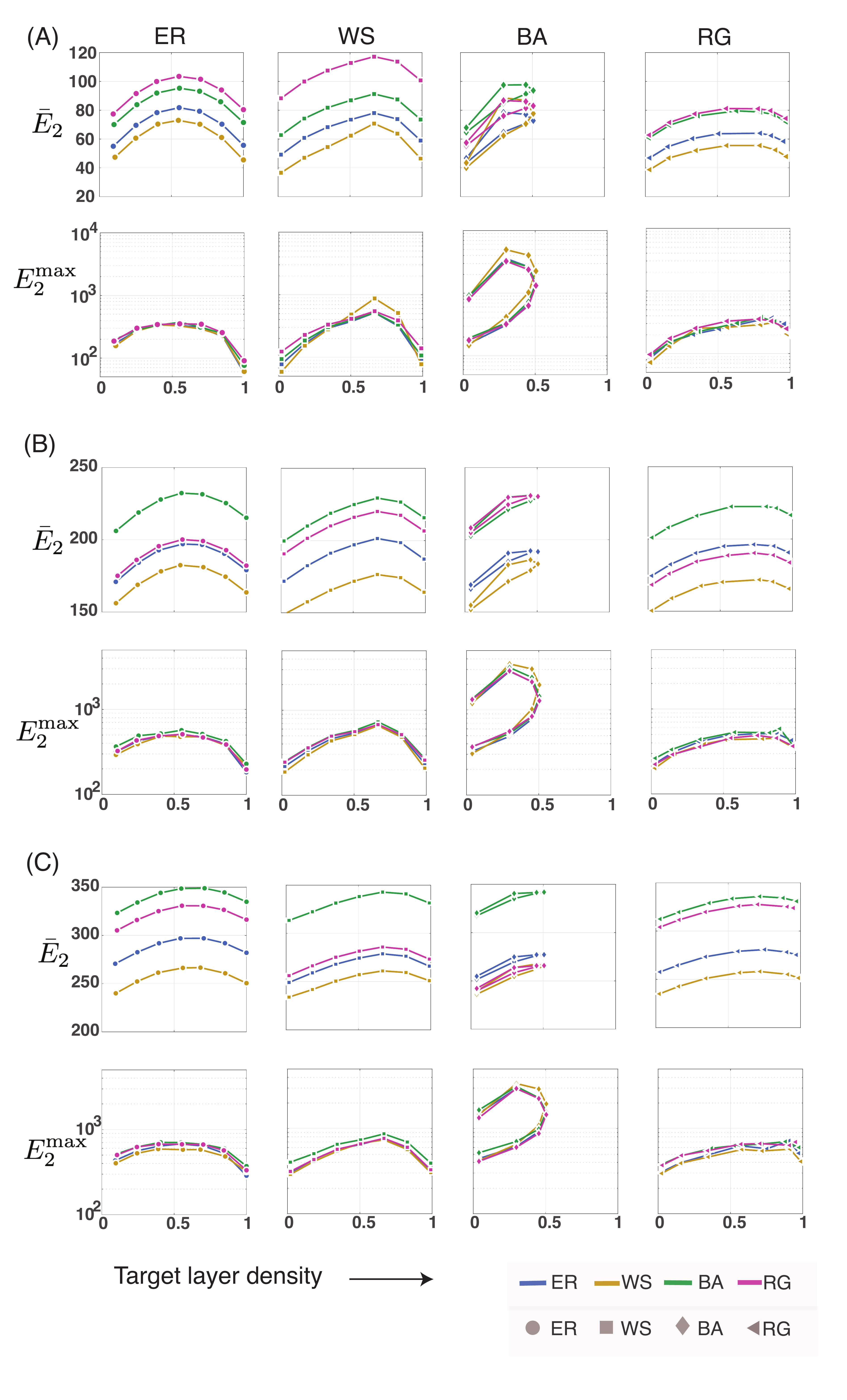}
    \caption{\textbf{Dependence of the target-layer control on the input-layer density.} Average control energy $\bar{E}_2$ and maximum control energy $E_2^{\rm{max}}$ plotted against the density of the target layer for three values of the input layer density: (A) $\rho_1 = 0.2$,
    (B) $\rho_1 = 0.4$, and (C) $\rho_1 = 0.5$. }
    \label{fig:fig6_rho1plots}
\end{figure}

The results in Section {\color{red} \bf III A}of the main text have been
obtained for a fixed density of the input layer. Here we show that the density
and topology dependence of target-layer control remain unchanged for different
densities of the input layer. Similar to the results shown in Section
{\color{red} \bf III A} of the main text, here we investigate the average and
maximum control energies for all 16 duplexes for three different values of the
input-layer density (Fig. \ref{fig:fig6_rho1plots}). We find that the results
remain unchanged.

\section{Control properties of the input layer}
\label{Ssec7:input_layer_control}
\begin{figure}[h]
    \centering
    \includegraphics[scale=0.35]{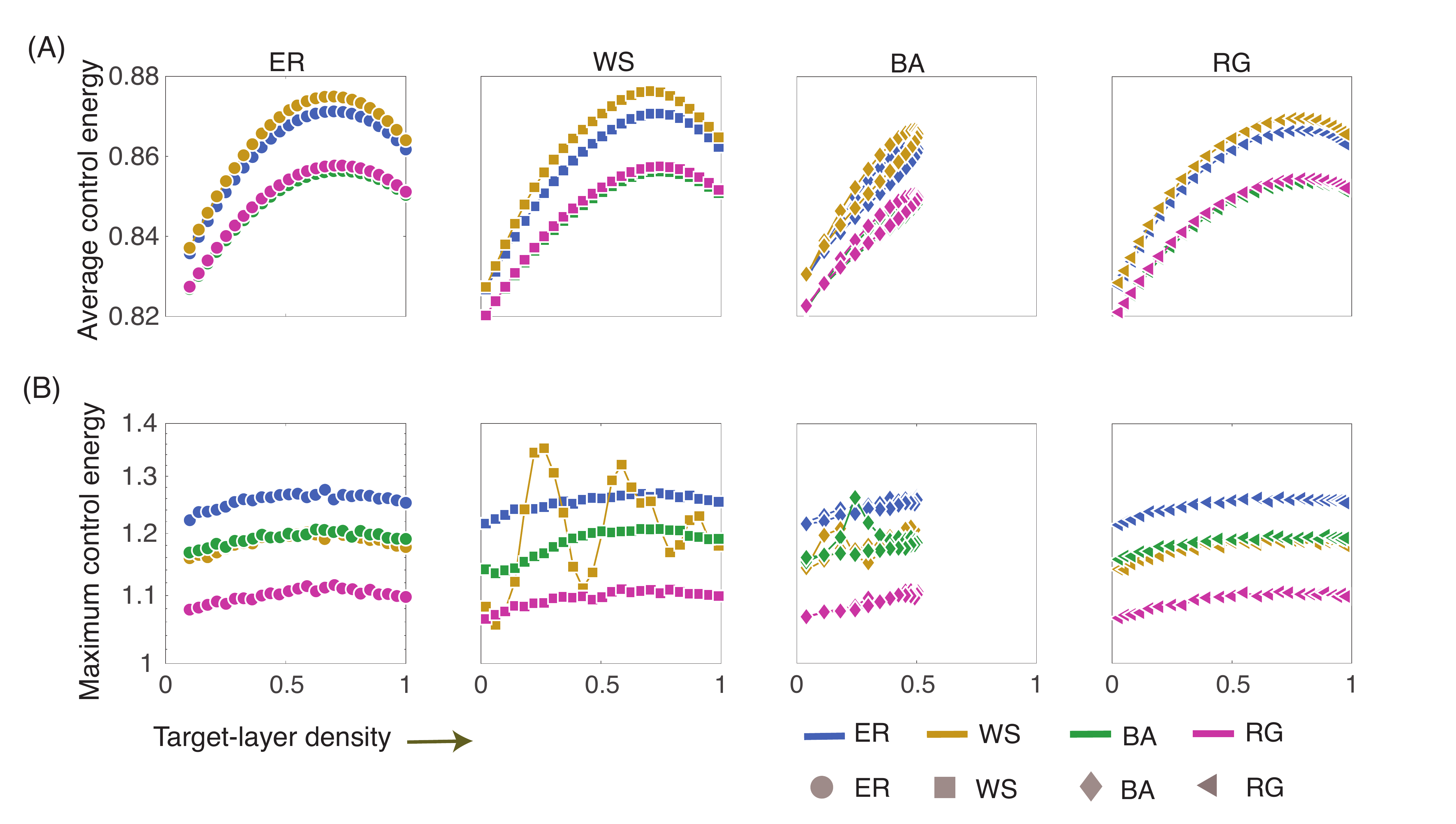}
    \caption{\textbf{Control properties of the input layer.} (A) Average control energy and (B) maximum control energy for the input layer plotted against the density of the target layer. These quantities are calculated for the same duplex networks as those in Fig. 2 of main text. Each column corresponds to the topology of the target layer as indicated at the top of panel. Differently-colored curves indicate the topology of the input layer as noted in the legend. }
    \label{fig:fig7_Layer1control}
\end{figure}

The average control energy and maximum control energy can also be examined for the input layer. We find a qualitatively similar dependence on the target-layer density for both quantities; the intermediate densities are consistently the hardest to control. An exception to this trend is the maximum energy curve for the WS-WS network (the yellow curve in the second plot of 
the bottom panel in Fig. \ref{fig:fig7_Layer1control}).

\section{Rotation in N-Dimensions}
\label{Ssec8:rotation}
In order to study the effect of alignment between the eigenmodes of the two layers of a duplex network, we construct a rotation matrix in $N$-dimensions, where $N$ is the number of nodes in each layer. In the main text, we calculate the cost of controlling the state of the second layer along its most unstable eigenmode (i.e. the eigenmode with the most positive eigenvalue) as a function of its alignment with the eigenmodes of the input layer. To perform this calculation, we choose the rotation plane consisting of the first two eigenmodes of the input layer denoted by $\mathbf{p}_1$ and $\mathbf{p}_2$, and we define the outer product $ \mathbf{B} = \mathbf{p}_1 \wedge \mathbf{p}_2$. The quantity $\mathbf{B}$ is a bivector that defines the plane containing the two eigenvectors $\mathbf{p}_1$ and $\mathbf{p}_2$, and is given as \cite{ChrisDoran_GeomAlgeb}: 

\begin{align}
    \mathbf{B} &= \underset{ij}{\Sigma} \epsilon_{ij} ( p_{1i} p_{2j} - p_{1j} p_{2i}) 
    \mathbf{e}_i \wedge \mathbf{e}_j, 
    \label{bivector}
\end{align}

\noindent 
Here, $\mathbf{e}_i$ and $\mathbf{e}_j$ denote external axes, and $\epsilon_{ij}$ 
is Levi-Civita symbol in two dimensions such that, 

\begin{align}
    \epsilon_{ij} &= 1, \quad \rm{if}~(i,j) = (1,2) \nonumber \\
    & = -1, \quad \rm{if}~(i,j) = (2,1) \nonumber\\
    &= 0, \quad \rm{if}~ i =j . 
\end{align}

The bivector $\mathbf{B}$ can be used to construct a rotor $R(\phi) = \exp(\mathbf{B} \phi/2)$ \cite{ChrisDoran_GeomAlgeb}. The equivalent rotation matrix $\mathcal{R}(\theta)$ is then given by $\mathcal{R}(\phi) = \exp(\phi M)$. Here, $M$ is a skew-symmetric matrix such that $M_{ij} = B_{ij}, \rm{for} \quad i<j$ and $M = -M^T$. We set $\phi = s \pi/2$, 
and vary $s$ in the range $[0, 1]$ to alter the alignment of the dominant eigenvector of the second layer in the plane $\mathbf{p}_1 \wedge \mathbf{p}_2$. 

\section{Optimal energy in the single-mode approximation}
\label{Ssec9:single_mode}
In this section, we obtain the solutions for state variables and optimal
control energy values in the one-mode limit. We begin from the effective state
equations in the one-mode limit (Eqs.~{\color{red} (10a)} and {\color{red} (10b)}
described in Section {\color{red} \bf IV} of the main paper):

\begin{align}
    \dot{w}_\alpha &= \xi_\alpha w_\alpha +  U_\alpha 
    \label{Seq:weq_1d}\\
    \dot{v}_\beta  &= \mu_\beta v_\beta + \kappa C_{\beta \alpha} w_\alpha .
    \label{Seq:veq_1d}
\end{align}

\noindent Here, $C_{\beta \alpha} = \cos \theta_{\beta \alpha}$. The underlying
assumption used to arrive at the above equations is that the structure and
behaviour of each layer is approximated by a single eigenmode: the
$\alpha^{\rm{th}}$ eigenmode in the input layer and the $\beta^{\rm{th}}$
eigenmode in the target layer. In the state equations for the full system,
$w_\alpha$ and $v_\beta$ denoted the weight of the state vector along the
retained eigenmodes, while $U_\alpha$ denoted the projection of the control
signal $\bf{u}$ along the $\alpha^{\rm{th}}$ eigenmode of the input layer (Eqs.~{\color{red} (5a)} 
and {\color{red} (5b)} of the main paper. 
In the one-mode limit, the direction of the state vector of each layer is parallel to the retained eigenmode with magnitudes $w_\alpha$ and $v_\beta$. The problem of finding the $U_\alpha(t)$ that minimizes the energy $E_\alpha = \int_0^{t_F} U_\alpha(t)^2 dt$ to navigate the system to a final state $(w_{\alpha F}, v_{beta F})$ at time $t= t_F$ from the initial state at origin can be stated in a similar way as Eq.~{\color{red}(6)} of the main paper:

\begin{align}
\underset{U_\alpha(t)}{{\rm{minimize}}} \quad E_\alpha \nonumber\\
\text{subject to }  \dot{w}_\alpha &= \xi_\alpha w_\alpha + U_\alpha \nonumber \\
\dot{v}_\beta &= \mu_\beta v_\beta + \kappa w_\alpha C_{\beta \alpha} .
\label{Seq:optim_statement1d} 
\end{align}

To solve this problem, we follow the approach outlined in Section \ref{Ssec1:exact_sol}. Denoting the co-state variables
corresponding to $w_\alpha$ and $v_\beta$ by $\lambda_\alpha$
and $\sigma_\beta$, the cost function takes the following form: 

\begin{align}
     J = \int_o^{t_F} (H (w_\alpha, v_\beta, U_\alpha) - \lambda_\alpha \dot{w}_\alpha 
     - \sigma_\beta \dot{v}_\beta)   dt, 
\label{J_1d}
\end{align}

\noindent where, the Hamiltonian $H$ is given by: 

\begin{align}
    H(w_\alpha, v_\beta, U_\alpha) = U_\alpha^2 + \lambda_\alpha (\xi_\alpha w_\alpha + U_\alpha ) + \sigma_\beta ( \mu_\beta v_\alpha + \kappa w_\alpha C_{\beta \alpha}) .
    \label{Seq:H_1d}
\end{align}

\noindent The dynamic equations for the co-state variables are then given by: 

\begin{align}
    \dot{\lambda}_\alpha &= - \frac{ \partial H}{\partial w_\alpha} \nonumber \\
    & = - \xi_\alpha \lambda_\alpha - \kappa \sigma_\beta C_{\beta \alpha}
    \label{Seq:lam_eq1d}\\
    \dot{\sigma}_\beta & = - \frac{ \partial H}{\partial v_\beta} \nonumber \\
    & = -\sigma_\beta \mu_\beta .
    \label{Seq:sig_eq1d}
\end{align}

Eqs.~\eqref{Seq:weq_1d}, \eqref{Seq:veq_1d}, \eqref{Seq:lam_eq1d}, and \eqref{Seq:sig_eq1d}
can be solved in combination with the boundary conditions: $\{w_{\alpha}(0), v_\beta(0) \} = \{ 0,0 \}$ and $\{w_{\alpha}(t_F), v_\beta(t_F) \} = \{ w_{\alpha F}, v_{\beta F} \}$. The solutions for each of these quantities can be written as follows: 

\begin{align}
     \sigma_\beta(t) & = \frac{1}{ \kappa^2 C_{\beta \alpha}^2 F} 
     \left[ G_{\sigma w} w_{\alpha F} + G_{\sigma v} v_{\beta F} \right],
     \label{Seq:sig1d_sol}\\
     \lambda_\alpha(t) &= \frac{1}{ \kappa C_{\beta \alpha} e^{t (\mu_\beta + \xi_\alpha)} F} 
     \left[ G_{\lambda w} w_{\alpha F} + G_{\lambda v} v_{\beta F}  \right], 
     \label{Seq:lam1d_sol}\\  
      w_\alpha(t) &=  \frac{1}{ \kappa C_{\beta \alpha} e^{t (\mu_\beta + \xi_\alpha)} F} 
         \left[ G_{w w} w_{\alpha F} + G_{w v} v_{\beta F}  \right],
         \label{eq:w1d_sol} \\ 
     v_\beta(t) &= \frac{1}{F}\left[ G_{vw} w_{\alpha F} + G_{vv} v_{\beta F} \right], 
     \label{Seq:v1d_sol}
\end{align}

\noindent
\noindent where the factor $F$ and the coefficients of $w_{\alpha F}$ and $v_{\beta F}$ in Eqs. \Crefrange{Seq:sig1d_sol}{Seq:v1d_sol} are given by: 

\begin{align}
    F &= \left[(e ^{2 \mu_\beta t_F} -1 ) (e^{2 \xi_\alpha t_F} -1 ) \right] (\mu_\beta^2 + \xi_\alpha^2) 
	 - 2 \left[ 1 + e^{2 \mu_\beta t_F} + e^{2 \xi_\alpha t_F} - 4 e^{(\mu_\beta + \xi_\alpha) t_F} + e^{2 (\mu_\beta + \xi_\alpha) t_F } \right] \mu_\beta \xi_\alpha 
	 \label{Seq:F}
\end{align}

\begin{align}
    G_{\sigma w} & = -4 \kappa C_{\alpha \beta} e^{(-t + t_F) \mu_\beta} \mu_\beta (\mu_\beta^2 - \xi_\alpha^2)  \left( ( e^{2 \xi_\alpha t_F } -1) \mu_\beta + (1 + e^{2 \xi_\alpha t_F } - 2 e^{t_F (\mu_\beta + \xi_\alpha)} \xi_\alpha) 
     \right), 
     \label{Seq:G_sigw}\\ 
    G_{\sigma v} & = - 4 e^{\mu_\beta( -t + t_F)} \mu_\beta ( e^{2 t_F \xi_\alpha} -1) ( \mu_\beta^2 - \xi_\alpha^2)^2, 
    \label{Seq:G_sigv}\\
    G_{\lambda w}  & = 4 \kappa C_{\beta \alpha} (\mu_\beta + \xi_\alpha )
                        [ e^{t_F \mu_\beta + \xi_\alpha t} \mu_\beta (\mu_\beta -\xi_\alpha)
			              + 2 \left( e^{ (t+t_F) \mu_\beta + 2 t_F \xi_\alpha } 
		               +  e^{2 t_F \mu_\beta + ( t+ t_F) \xi_\alpha } \right) \mu_\beta \xi_\alpha \nonumber\\
			         & - e^{ t_F \xi_\alpha + t \mu_\beta} \xi_\alpha (\mu_\beta-\xi_\alpha)
			         - e^{ t \xi_\alpha + t_F (\mu_\beta + 2 \xi_\alpha) }\mu_\beta (\mu_\beta + \xi_\alpha)
			      - e^{ t \mu_\beta + t_F (\xi_\alpha + 2 \mu_\beta) } \xi_\alpha (\mu_\beta+\xi_\alpha) ] 
    \label{Seq:G_lamw}\\
	G_{\lambda v} &= 4\mu_\beta ( \mu_\beta^2 - \xi_\alpha^2 ) \left[- 2 e^{t\mu_\beta                  + t_F\xi_\alpha}\xi_\alpha 
				     + 2 e^{ (t+t_F)\mu_\beta + 2t_F\xi_\alpha}\xi_\alpha 
				    + e^{ \mu_\beta t_F + \xi_\alpha t }( \mu_\beta +\xi_\alpha )
				     - e^{ \xi_\alpha t + ( \mu_\beta + 2\xi_\alpha) t_F } (\mu_\beta + \xi_\alpha) \right]
	 \label{Seq:G_lamv}\\
	G_{ww} &= \kappa 
	          C_{\beta \alpha} [ ( 2 e^{ t_F \mu_\beta + \xi_\alpha t} - 2 e^{ \mu_\beta (t+ t_F) + 2 \xi_\alpha t} ) 
	                  \mu_\beta ( \mu_\beta-\xi_\alpha )
                   + e^{ \mu_\beta(t + 2 t_F)\mu_\beta + \xi_\alpha (2 t + t_F) }
                      (\mu_\beta - \xi_\alpha)^2 \nonumber \\ 
            &       + 4 e^{ 2\mu_\beta t_F + \xi_\alpha (t + t_F) } \mu_\beta \xi_\alpha
		            + 2 e^{\mu_\beta t_F  + 2 \xi_\alpha t_F }
		            (  e^{\mu_\beta t} - e^{\xi_\alpha t} ) 
		            \mu_\beta (\mu_\beta + \xi_\alpha) \nonumber\\
            &       + e^{\mu_\beta t + 2 \xi_\alpha t + \xi_\alpha t_F }
                   ( \mu_\beta^2-\xi_\alpha^2 )
                   - e^{\mu_\beta ( t + 2 t_F )+ \xi_\alpha t_F}
                   ( \mu_\beta + \xi_\alpha )^2
                + e^{\mu_\beta t +\xi_\alpha t_F} (\xi_\alpha^2- \mu_\beta^2) ] 
     \label{Seq:G_ww}\\
     G_{wv} &= -2 \mu_\beta (\mu_\beta^2- \xi_\alpha^2) 
                   \left[ e^{(\mu_\beta + 2 \xi_\alpha) t } 
                   \left( e^{\mu_\beta t_F} - e^{\xi_\alpha t_F} \right) 
                    + e^{\mu_\beta t_F + \xi_\alpha t } (e^{2 \xi_\alpha t_F} -1)
                                - e^{\mu_\beta t + \xi_\alpha t_F}(e^{(\mu_\beta + \xi_\alpha) t_F} -1 ) \right]
    \label{Seq:G_wv}\\
    G_{vw} & = 4 \kappa C_{\beta \alpha} e^{(\mu_\beta + \xi_\alpha) t_F} 
               [ \cosh(\mu_\beta t_F) \left( \xi_\alpha \sinh(\mu_\beta t) 
                       - \mu_\beta \sinh(\xi_\alpha t) \right) \nonumber \\ 
            &           + \cosh(\xi_\alpha t_F) \left(-\xi_\alpha \sinh(\mu_\beta t) 
                       + \mu_\beta \sinh(\xi_\alpha t) \right) \nonumber \\
            &          + (\cosh(\mu_\beta t) - \cosh(\xi_\alpha t)) (-\xi_\alpha \sinh(\mu_\beta t_F) + \mu_\beta \sinh(\xi_\beta t_F)) ]
    \label{eq:G_vw} \\
    G_{vv} & = 4 e^{(\mu_\beta + \xi_\alpha) t_F} 
                    [ \mu_\beta \left(-\cosh(\mu_\beta t_F) + \cosh(\xi_\alpha t_F)
                     + \sinh(\mu_\beta t_F) \right) \nonumber \\
            &       \left(- \xi_\alpha (\cosh(\mu_\beta t) - \cosh(\xi_\alpha t) 
                     + \sinh(\mu_\beta t)) + \mu_\beta \sinh(\xi_\alpha t) \right) \nonumber \\
            &         + ( \mu_\beta^2 \cosh(\mu_\beta t) + 
                            \xi_\alpha^2 \sinh(\mu_\beta t) 
                            - \mu_\beta ( \mu_\beta \cosh(\xi_\alpha t) 
                            + \xi_\alpha \sinh(\xi_\alpha t) ) )\sinh(\xi_\alpha t_F) ] .
    \label{Seq:G_vv}
\end{align}

Finally, the optimal control signal $U_\alpha$ in the one-mode limit is determined from $\lambda_\alpha(t)$ using the relation $U_\alpha(t) = -\frac{\lambda_\alpha(t)}{2}$. Following this formulation, the optimal control energy $E_\alpha$ can be calculated as: 

\begin{align}
    E^\ast_\alpha = \frac{1}{Z} \left[ E_{ww} w_{\alpha F}^2 + E_{vv} v_{\beta F}^2 + E_{wv} w_{\alpha F} v_{\beta F}  \right]. 
    \label{Seq:E_alpha}
\end{align}

\noindent Here the coefficients $E_{ww}, E_{wv}, E_{wv}$ and the denominator $Z$ are given by: 

\begin{align}
    E_{ww} &= 2 \kappa^2 C_{\beta \alpha}^2 (\mu_\beta + \xi_\alpha) 
    \left[  (e^{2 \xi_\alpha t_F} -1 ) \mu_\beta^2 
            + (e^{2 \mu_\beta t_F} -1 ) \xi_\alpha^2 
            + (2 + e^{2 \mu_\beta t_F} + e^{2 \xi_\alpha t_F}  
            - 4 e^{(\mu_\beta +\xi_\alpha) t_F}) \mu_\beta \xi_\alpha \right] 
    \label{Seq:E_ww} \\
    E_{vv} & = 2 \mu_\beta (e^{2 \xi_\alpha t_F} - 1 ) 
              \left( \mu_\beta^2 - \xi_\alpha^2 \right)^2
    \label{Seq:E_vv} \\
    E_{wv} & = 4 \kappa C_{\beta \alpha}
               \mu_\beta \left( \mu_\beta^2 - \xi_\alpha^2 \right)
               \left( (e^{2 \xi_\alpha t_F} - 1) \mu_\beta 
               + (1  + e^{2 \xi_\alpha t_F} -2 e^{(\mu_\beta + \xi_\alpha) t_F}) \xi_\alpha \right)
    \label{Seq:E_wv}\\
     Z &= \kappa^2 C_{\beta \alpha}^2 \left[(e ^{2 \mu_\beta t_F} -1 ) (e^{2 \xi_\alpha t_F} -1 ) \right] (\mu_\beta^2 + \xi_\alpha^2) 
	 - 2 \left[ 1 + e^{2 \mu_\beta t_F} + e^{2 \xi_\alpha t_F} - 4 e^{(\mu_\beta + \xi_\alpha) t_F} + e^{2 (\mu_\beta + \xi_\alpha) t_F } \right] .
	 \label{Seq:Z}
\end{align}

\noindent Taking Eqs.~\Crefrange{Seq:E_alpha}{Seq:Z} together, we obtain the expression 
$E^\ast_\alpha = S_{-2} C_{\beta \alpha}^{-2} + S_{-1} C_{\beta \alpha}^{-1} + 
S_0$ discussed in Section {\color{red} \bf IV} of the main paper of the main paper. \\

In addition to the final state $\{w_{\alpha F}, v_{\beta F}\}$, the energy $E_\alpha$ depends on $\xi_\alpha, \mu_\beta$ and the parameter $C_{\beta \alpha}$.
The dependencies on the ratio $\mu_\beta/\xi_\alpha$ and $C_{\beta \alpha}$ are plotted
in Fig.~{\color{red} 4}(B)-(C) of the main paper. Here, we plot the dependence on 
$\xi_\alpha$ for the state $\{w_{\alpha F}, v_{\beta F} \} = \{1,0\}$ in Fig.~\ref{fig:fig9_E_vs_xi}.

\section{Association between optimal control energies and eigenvalues}
\label{Ssec10:E_vs_eigval}
In this section, we seek to determine whether there is an association between the optimal control energies (to move the state of each layer of the duplex networks along their eigenvectors by a unit amount) and the corresponding eigenvalues of those eigenvectors. Accordingly, we examine the optimal control energies as a function of the eigenvalues (Fig. \ref{fig:fig10_E_vs_eigs}). In these numerical assessments, we observe a trend similar to that predicted by the formal analysis in the single-mode limit.

\newpage

\begin{figure}
    \centering
    \includegraphics[scale=0.35]{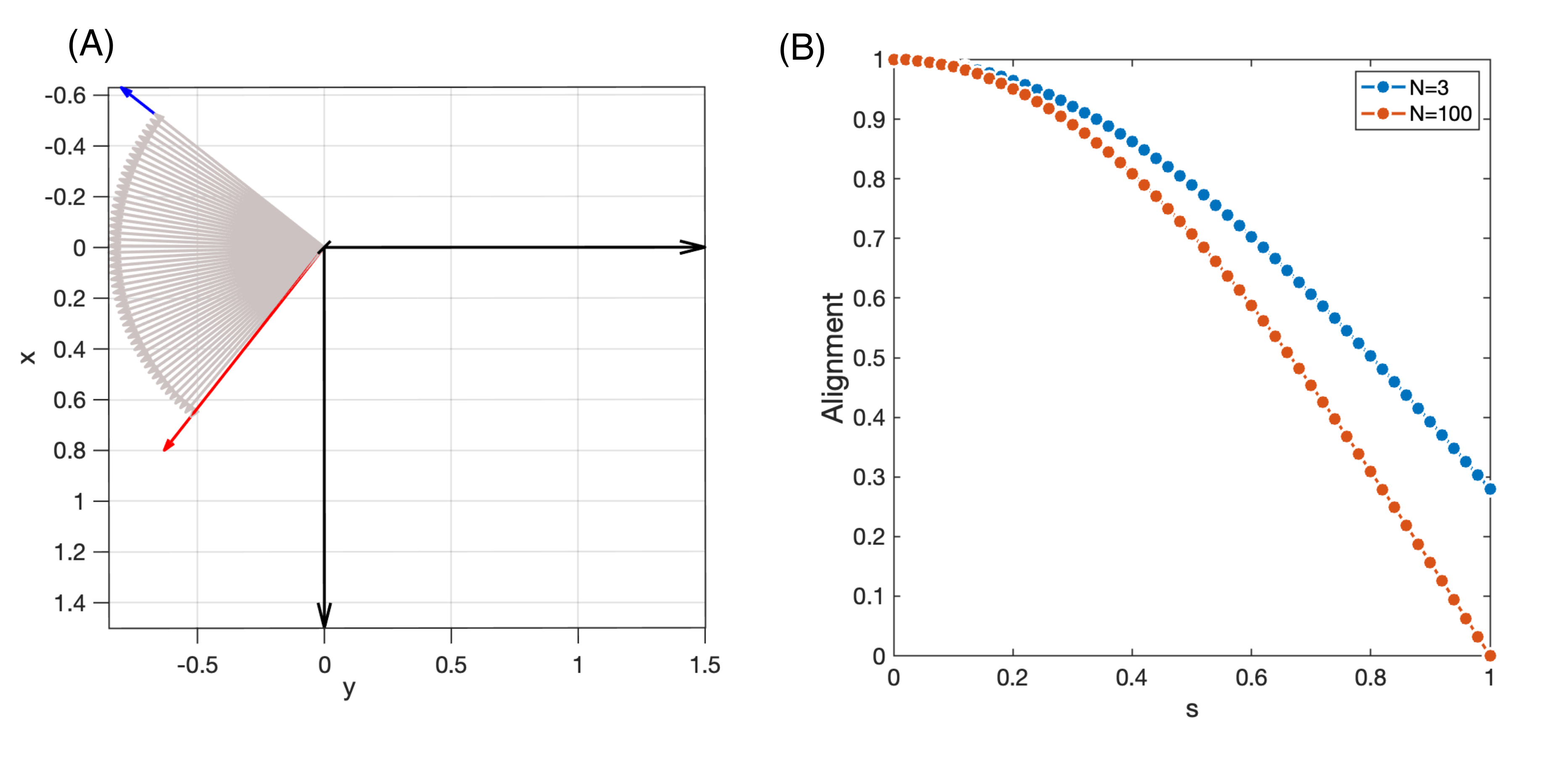}
    \caption{\textbf{Rotation in a plane.} (A) Rotation of a random initial vector in 3-dimensions (blue arrow) by the rotation matrix $\mathcal{R}(s)$ in $x-y$ plane. (B) Alignment between a rotated vector and an initial vector as a function of the parameter $s$ for $N=3$ and for $N=100$. For $N=100$, the initial vector is in the plane of rotation; the component of the initial vector perpendicular to the rotation plane is zero. This relation results in spanning the complete alignment range (from initially perfect to orthogonal) as a function of $s$ for $N=100$.}
    \label{fig:fig8_rotation}
\end{figure}

\begin{figure}
    \centering
    \includegraphics[scale=0.5]{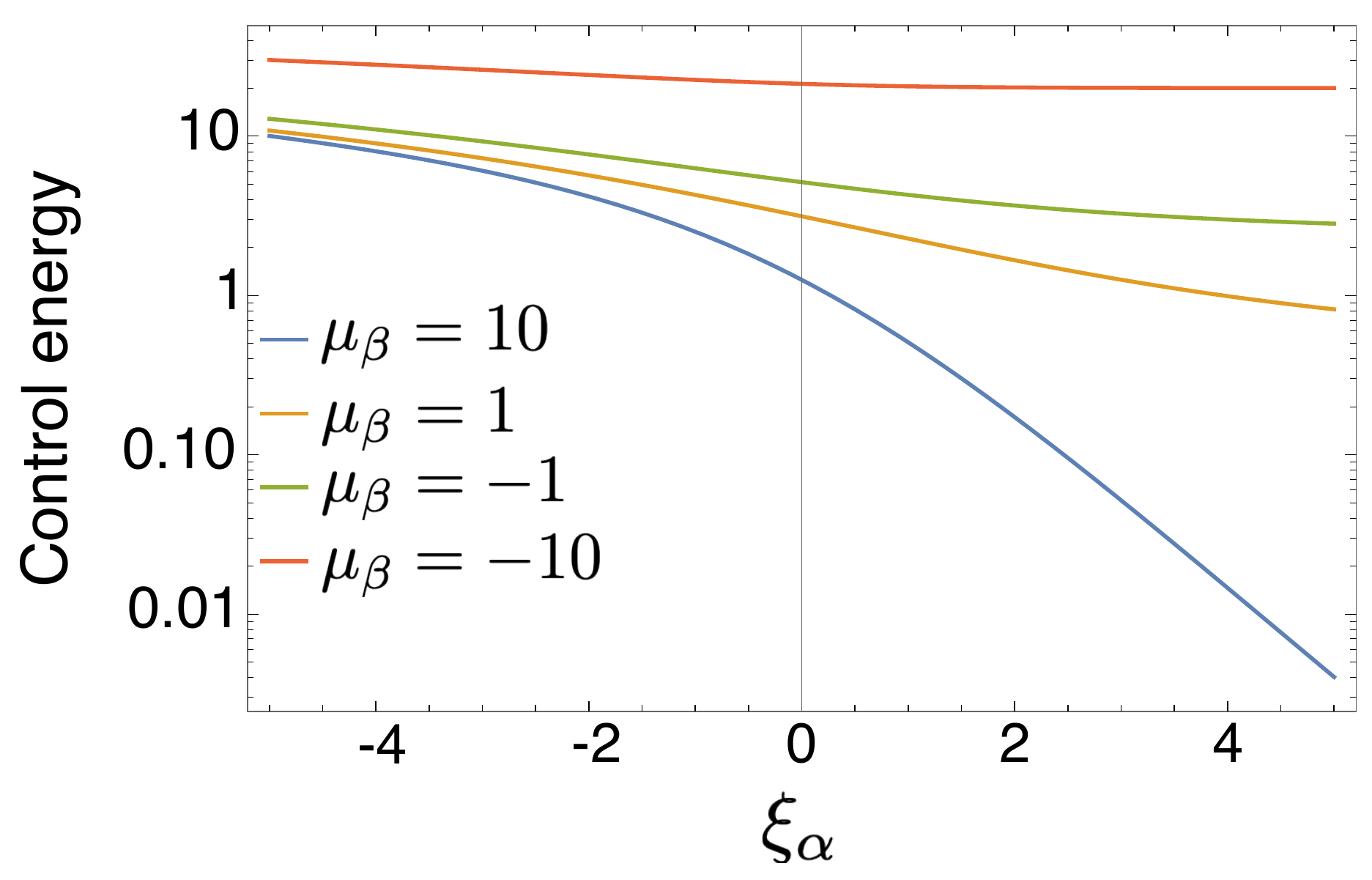}
    \caption{\textbf{Dependence of control energy on the eigenvalue $\xi_\alpha$ in the one-mode limit.} The control energy $E_\alpha$ corresponding to the final state $\{w_{\alpha F}, v_{\beta F} \} = \{1,0\}$ is plotted as a function of $\xi_\alpha$ for 4 different values of $\mu_\beta$ in the one-mode limit. These curves confirm that a faster decaying mode (a more negative eigenvalue) has higher control cost, while a faster growing mode (a more positive eigenvalue) has a lower control cost.}
    \label{fig:fig9_E_vs_xi}
\end{figure}

\begin{figure}
    \centering
    \includegraphics[scale=0.3]{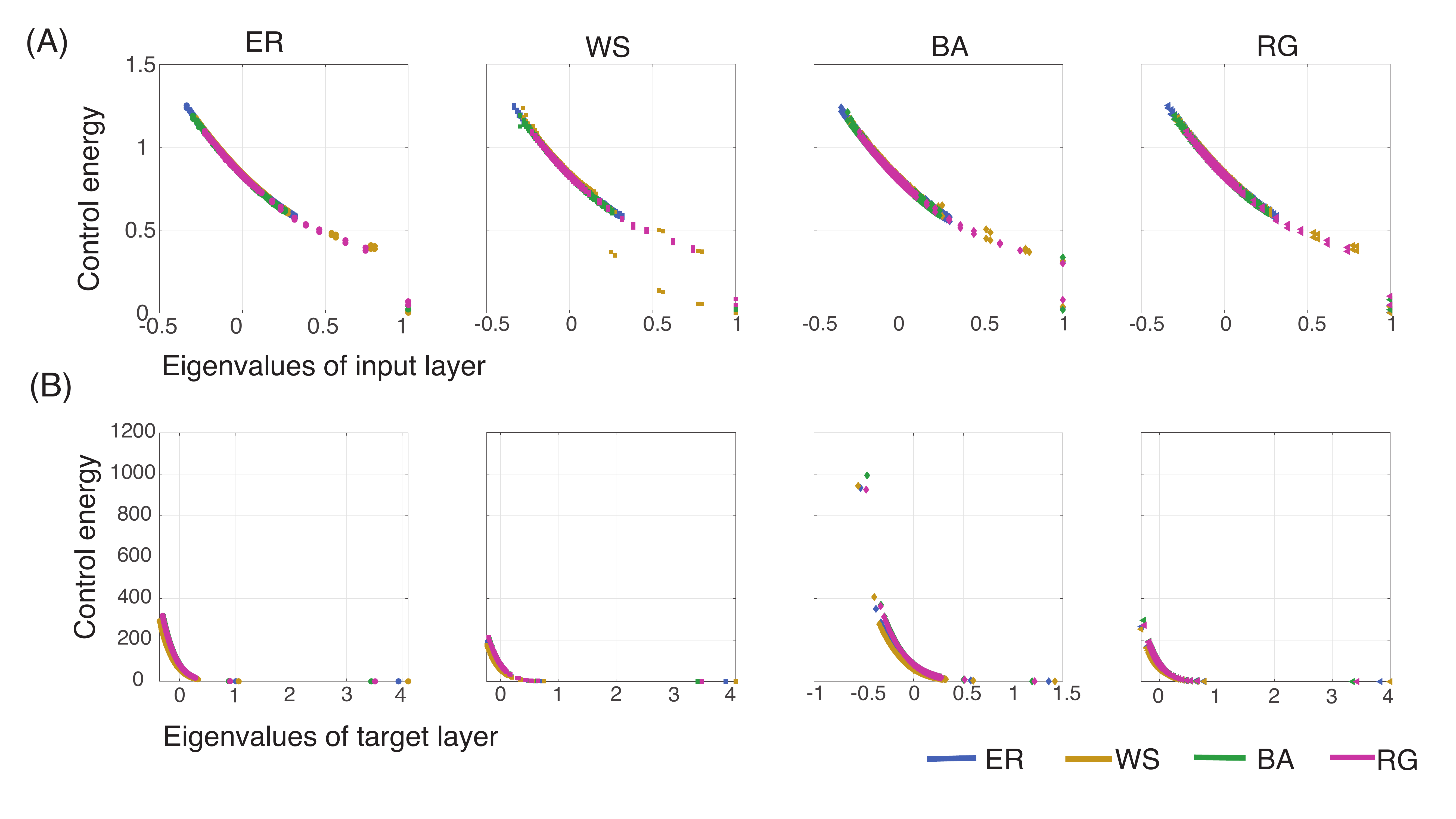}
    \caption{\textbf{Association between optimal control energies and eigenvalues.} Optimal control energies to move the state of (A) the input layer and (B) the target layer along their eigendirections plotted against their respective eigenvalues. The curves show a strong association between optimal control energies and the corresponding eigenvalues, similar to that predicted by the analytical results of the single-mode limit. Each column of the figure corresponds to the topology of the target layer as indicated on the top of the panel. Within each plot, different colors correspond to the topology of the input layer. 
    }
    \label{fig:fig10_E_vs_eigs}
\end{figure}

\bibliography{Refs_multicontrol_theory}
\bibliographystyle{ieeetr}